# Picosecond time-resolved antibunching measures nanoscale exciton motion, annihilation, and true number of chromophores

*Gordon J. Hedley[1,†,*], Tim Schröder[2,†], Florian Steiner[2], Theresa Eder[3], Felix Hofmann[3], Sebastian Bange[3], Dirk Laux,[4] Sigurd Höger,[4] Philip Tinnefeld[2], John M. Lupton[3] and Jan Vogelsang[3,*]*

[1]School of Chemistry, University of Glasgow, University Avenue, Glasgow, G12 8QQ, United Kingdom

[2]Department Chemie and Center for NanoScience (CeNS), Ludwig-Maximilians-Universität München, Butenandtstraße 5-13 Haus E, 81377 München, Germany

[3]Institut für Experimentelle und Angewandte Physik, Universität Regensburg, Universitätsstraße 31, 93053 Regensburg, Germany

[4]Kekulé-Institut für Organische Chemie und Biochemie, Universität Bonn, Gerhard-Domagk-Strasse 1, 53121 Bonn, Germany

[†] These authors contributed equally to the work

* Gordon.Hedley@glasgow.ac.uk

* Jan.Vogelsang@physik.uni-regensburg.de

## Abstract

The particle-like nature of light becomes evident in the photon statistics of fluorescence of single quantum systems as photon antibunching. In multichromophoric systems, exciton diffusion and subsequent annihilation occurs. These processes also yield photon antibunching but cannot be interpreted reliably. Here, we develop picosecond time-resolved antibunching (psTRAB) to identify and decode such processes. We use psTRAB to measure the true number of chromophores on well-defined multichromophoric DNA-origami structures, and precisely determine the distance-dependent rates of annihilation between excitons. Further, psTRAB allows us to measure exciton diffusion in mesoscopic H- and J-type conjugated-polymer aggregates. We distinguish between one-dimensional intra-chain and three-dimensional inter-chain exciton diffusion at different times after excitation and determine the disorder-dependent diffusion lengths. Our method provides a new lens through which excitons can be studied at the single-particle level, enabling the rational design of improved excitonic probes such as ultra-bright fluorescent nanoparticles, and materials for optoelectronic devices.

## Introduction

In a wide range of fluorescent nanoparticles such as conjugated polymers, semiconductor quantum dots, perovskite nanoparticles, light-harvesting complexes and many other natural or synthetic multichromophoric nanoparticles (mcNP), multiple excitons can exist simultaneously and in close proximity to each other.[1-10] The number of chromophores as well as their interactions through exciton diffusion and annihilation processes are key parameters to describe the photophysical characteristics of mcNPs such as brightness,[11, 12] photoluminescence (PL) lifetime, exciton harvesting efficiency[13] and photostability,[12, 14] all of which are also important for the performance of materials in optoelectronic devices. Photon antibunching has been used to count chromophores,[15-17] however, this is typically not viable when exciton diffusion and singlet-singlet annihilation (SSA) occur as illustrated in Figure 1a. Single-photon emission from mcNPs has been interpreted as evidence of long-range interchromophore interactions in a number of large multichromophoric systems.[1, 2, 8, 18-21] However, in these cases information about the number of physical chromophores in the mcNPs is lost. Here, we demonstrate that picosecond time-resolved antibunching (psTRAB) can be used to disentangle information on the number of physical chromophores and exciton diffusion and annihilation processes. psTRAB exploits the fact that exciton diffusion and annihilation are time-dependent processes. Fingerprints of these processes are thus concealed in the PL photon stream of antibunching experiments under pulsed excitation.[6]

The degree of single photon emission is commonly measured by two photodetectors in a Hanbury Brown and Twiss (HBT) geometric configuration and is therefore sensitive to two-photon events. With this technique, it is either possible to count the number of chromophores, provided that SSA is neglected, or to measure the SSA rate if the exact number of chromophores is known. In practical situations, neither the number of chromophores nor the SSA rate are usually known for mcNPs, which severely limits the usefulness of this conventional technique.

With psTRAB, we analyse the photon stream of antibunching experiments with pulsed excitation by grouping photons with respect to their arrival time after the laser pulse and cross-correlating them to determine the probability of consecutive emission of two photons. Immediately after a laser pulse, SSA has not yet occurred and the emitted photons exhibit photon statistics corresponding to the number of physical chromophores present. As exciton diffusion and annihilation begin to dominate, the number of independent emitters decreases. Thus, the time-dependence of the photon statistics synchronised by the laser pulse reports on (i) the number of physical emitters present and (ii) the time evolution of exciton diffusion and annihilation.

To demonstrate the psTRAB technique, we have used DNA origami to construct mcNPs with a well-defined number of chromophores and well-defined spacing between them to accurately measure annihilation and benchmark our method. We then measure psTRAB of mesoscopic deterministic aggregates of conjugated polymers – the building blocks of films used in optoelectronic devices.[2] There we find that during the first 250 ps after excitation, diffusion of excitons mainly occurs between one and two dimensions, both along the polymer backbone and between π-stacked chains. The diffusion then becomes three-dimensional at later times, with an order-of-magnitude difference in the rate of annihilation between ordered H-type aggregates and disordered J-type aggregates. We can also extract the exciton diffusion lengths using the unique knowledge psTRAB gives on the number of independent chromophores present.

**Picosecond time-resolved antibunching (psTRAB)**

Our approach exploits the ability of modern time-correlated single-photon counting (TCSPC) hardware to record the absolute arrival time of a photon on each detector, both with respect to the start of the experiment, but also with respect to the last laser pulse (denoted as the microtime) as shown in Figure 1b. As an example, consider a nanoparticle with 5 physical chromophores as depicted by the black discs in Figure 1c. Absorption of a short pulse of light will create a Frenkel exciton (blue dot). The exciton can hop from one chromophore to another, e.g. by homo-FRET,[22] in a process referred to as exciton diffusion.[23] Now, if we consider the case where two excitons are created by the same excitation pulse, this hopping allows the excitons to move so that they become adjacent to each other and can annihilate by SSA.[19, 20] This process has a strong distance dependence due to the underlying FRET mechanism by which SSA occurs and is often hard to study in a quantitative manner.[24] By inspecting individual mcNPs on a confocal microscope with two single-photon detectors (Figure 1b) combined with TCSPC we measure the correlation events, $N$, dependent on the difference in photon arrival times, $\Delta t$, between photon events. We are thereby sensitive to the presence of two excitons in the mcNP. A histogram of $\Delta t$ delay times in integer units of the excitation-pulse period $T$ shows the number of photon detection coincidences from either one excitation pulse or from two separate excitation pulses (Figure 1c, right column). The ratio of the magnitude of the central peak at $\Delta t = 0$ to that of the lateral peaks, $N_\mathrm{c}/N_\mathrm{l}$, provides a measure for the number of independent chromophores, $n$, according to[16]

$$n = 1/(1 - \frac{N_\mathrm{c}}{N_\mathrm{l}}) \quad \text{eq. (1)}$$

By analysing the statistics of the PL photons detected at different time intervals after photoexcitation (panel c, second column), we can construct corresponding picosecond-resolved histograms of the photon statistics and thus measure how many independently emitting chromophores exist on a particular timescale. This is illustrated schematically in Figure 1c for a 5-chromophore mcNP. The left column depicts the evolution of randomised typical examples of such independent chromophores after a single laser excitation event, whereas the histograms in the middle and right columns are an accumulation of multiple excitation cycles to show the time-averaged result. At early times after excitation (panel c, first row), the two excitons contributing to $N_c$ events (blue dots) have had no time to interact or move via homo-FRET to neighbouring chromophores. From the photon coincidence histogram (right panel) we obtain a value of $n = 5$ with eq. 1. At a later time (panel c, second row), an exciton on a neighbouring physical chromophore may have, for example, interacted through SSA, and consequently excitation of such chromophores thus does not contribute to $N_c$ anymore, and we obtain $n = 4$ independent chromophores accordingly. These diffusion/SSA processes continue as a function of time, reducing the number of independent chromophores that could support the second exciton. Ultimately, at late times after the excitation pulse, only single photons can be detected because excitons on any other physical chromophore would have had enough time to diffuse and annihilate, yielding $N_c = 0$ and $n = 1$ (panel c, last row). This evolution of the photon statistics and the corresponding number of *independent* chromophores with time gives us a metric for the effective rate of exciton decay and provides direct microscopic insight into exciton annihilation and diffusion in mcNPs.

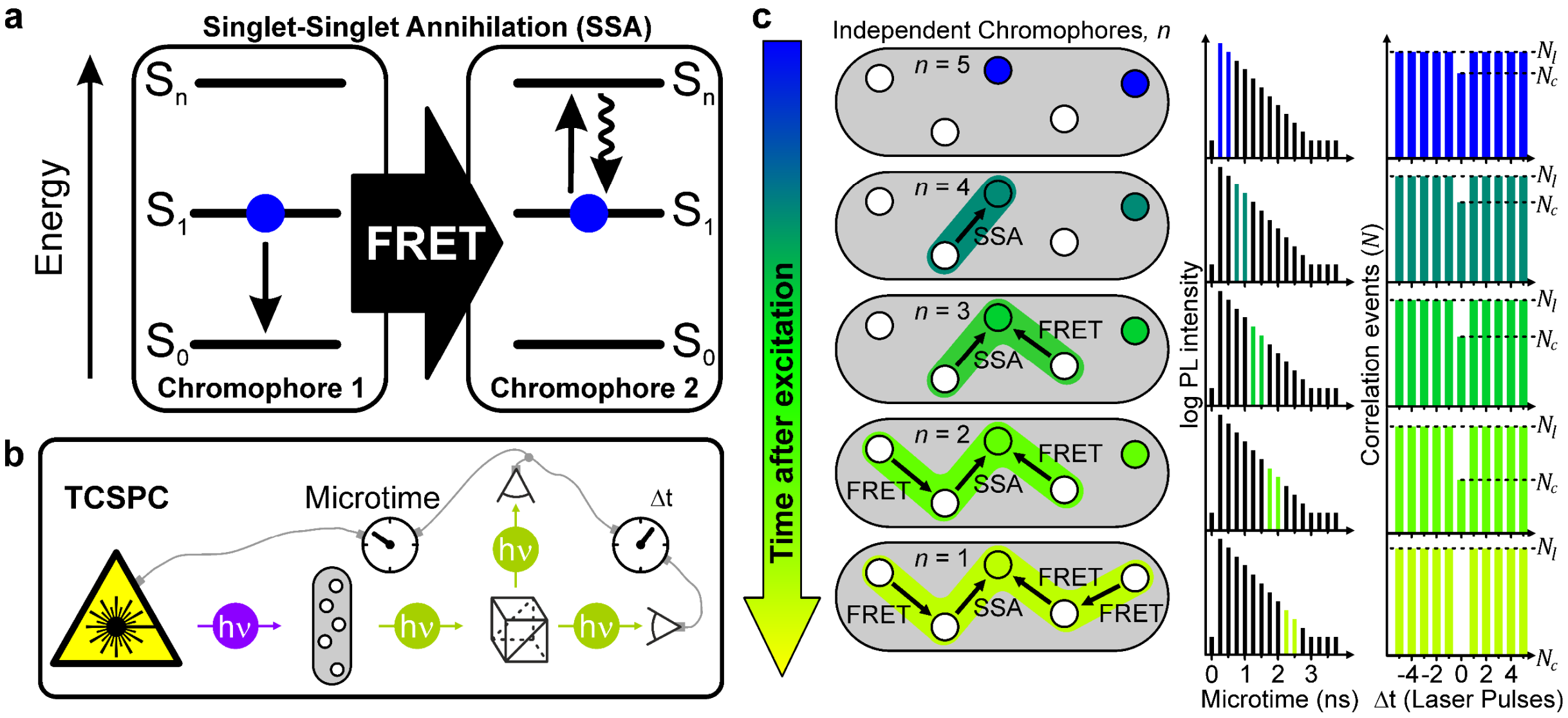

**Figure 1. Picosecond time-resolved antibunching (psTRAB).** (a) Singlet-Singlet annihilation (SSA) of a singlet exciton, $S_1$, on chromophore 1 by Förster resonance energy transfer (FRET) to an exciton on chromophore 2, which excites it into a higher excited state, $S_n$. Subsequently, chromophore 2 relaxes by internal conversion into its first excited state, $S_1$. Thus, the exciton (blue dot) on chromophore 1 is annihilated. (b) Principle of time-correlated single-photon counting (TCSPC) combined with a Hanbury Brown and Twiss photon correlator. A pulsed laser (purple) excites a multichromophoric nanoparticle (mcNP) (grey area). The statistics of the PL photon stream (green) is analysed by cross correlating two photon detectors. TCSPC yields the time difference between excitation and emission events, i.e. the "microtime", and the time difference between consecutive emitted photons, $\Delta t$, as determined by the repetition period of the pulsed laser. (c) On the left, 5 chromophores (discs) in a mcNP are shown schematically

with two singlet excitons (coloured discs), which after excitation can diffuse by site-to-site hopping, i.e. by homo-FRET and annihilate by SSA as a function of the excited-state lifetime. The overall PL decay, constructed from the microtimes, is shown in the centre, with the corresponding gated arrival time windows of photons used to construct the antibunching histograms stated in the righthand column. The ratio, $N_c/N_l$, of the number of correlation events in the central peak at $\Delta t = 0$, $N_c$, versus those in the lateral time-lagged peaks, $N_l$, allows us to determine the number of independent chromophores, $n$. As excitons diffuse through homo-FRET and annihilate through SSA, $n$ drops with time.

## Results

**Exciton annihilation in multichromophoric DNA origami nanoparticles.** To explore the fundamental nature of exciton diffusion and SSA it is desirable to have the best possible control over the number of dye molecules and their spatial position in the mcNP. The dyes need to be within distances to each other corresponding to the range of FRET of ~1-10 nm. We have therefore turned to the method of three-dimensional DNA origami to construct highly defined mcNPs. Similar structures have been used previously to study motor proteins and to characterize super-resolution microscopy techniques, and are modified here for our needs.[25, 26] The sketch in Figure 2a shows a short section of a 12-helix bundle with 6 inner and 6 outer helices. The total length of this DNA origami structure is ~225 nm. Five labelling positions separated by ~3 nm each are available in the centre of this modular structure.

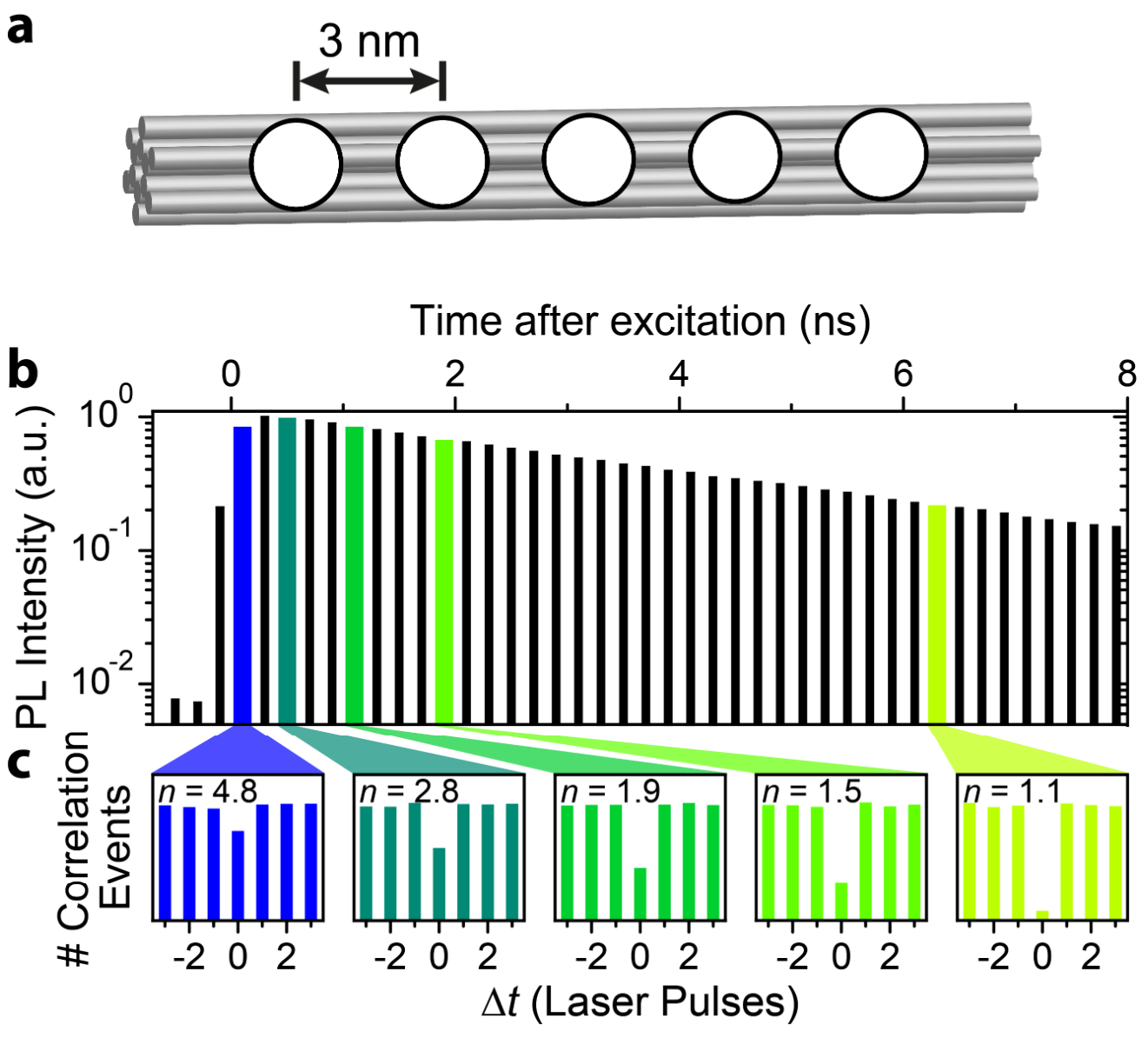

**Figure 2. Tracking exciton diffusion and annihilation in space and time on a well-defined multichromophoric DNA origami structure.** (a) Schematic of a short part of a 225 nm long 12-helix-bundle DNA origami structure with 6 inner and 6 outer helices. Five dyes (white discs) can be positioned at 3 nm spacing from each other. (b) Measured photoluminescence (PL) decay of a single DNA origami structure with five ATTO647N dye molecules. A single-exponential decay is observed with a PL lifetime of ~4.2 ns. Panel (c) shows five corresponding photon statistics histograms for different microtime bins (0-200 ps, 200-400 ps, 800-1000 ps, 1,600-1,800 ps and 6,400-6,600 ps) in terms of the photon correlation events, $N$, of the two photodetectors. Data were accumulated for 54 single mcNPs. The number of independent chromophores, $n$, determined from the correlation histogram for each microtime bin is stated in the histograms.

Based on this 12-helix bundle DNA origami structure, we designed seven different structures with different numbers of dyes and different distances between the dyes (see Supplementary Information for details of DNA origami structures). For the dye we chose ATTO647N, which is highly photostable and bright in the presence of a reducing and oxidizing system (ROXS).[27] The origami structures were examined on a custom-made confocal fluorescence microscope as described in the methods section (a typical PL transient is shown in Figure S3).[12] We begin discussing the mcNP with all five dye attachment positions filled with a dye. Figure 2b displays a histogram of photon arrival times, i.e. microtimes, in steps of 200 ps following pulsed excitation with a 640 nm laser. The PL decay is single exponential with a lifetime of 4.2 ns, which is typical for this dye attached to DNA and implies that no strong interchromophoric interactions occur.[12]

For this five-dye sample we select 200 ps time windows from the microtime histogram (coloured bars) and calculate the photon statistics for each gate as shown in Figure 2c. We used the peak of the instrument response function (see Figure S4) to determine zero microtime in the calculations. According to eq. 1, we estimate the number of independent chromophores, $n$, in the first 200 ps after excitation to be ~4.8, very close to the expected starting value of 5. Between 200 and 400 ps, $n$ drops to ~2.8 and reaches ~1.1 between 6,400 and 6,600 ps. The photons emitted by the five-chromophore structure at the latest times tend to be almost perfectly antibunched. In total, photon events of 54 individual mcNPs were accumulated to obtain enough correlation events for this analysis. The five histograms in Figure 2c reveal the timescale on which the excitons annihilate with each other to lower the number of independently emitting chromophores from five to one. We note that the fact that the number of chromophores inferred at the earliest times is slightly lower than the expected value of five can be explained by SSA having already occurred during the first 200 ps. One immediate conclusion of this method is that the number of dyes can be measured in an mcNP directly, even if the dyes are not emitting independently. Such knowledge is crucial in quantitative spectroscopic methods.[17, 28] A further crucial observation is that, in contrast to ensemble measurements,[29] the PL decay retains its monomolecular single-exponential form even though SSA clearly occurs. This is a particularly important observation because the non-exponentiality of ensemble PL decays, i.e. a bimolecular decay, is generally used to extract exciton encounter rates to infer diffusion lengths. In the ensemble, this approach only works at very high excitation fluences which are far from the population densities relevant to devices. However, it is crucial to realise that SSA always occurs, even at the lowest excitation fluences, because exciton diffusion always occurs. Our photon correlation technique is sensitive precisely and only to these rare events of double-chromophore excitation, which can be reached at arbitrarily weak fluences at the cost of extended integration times.

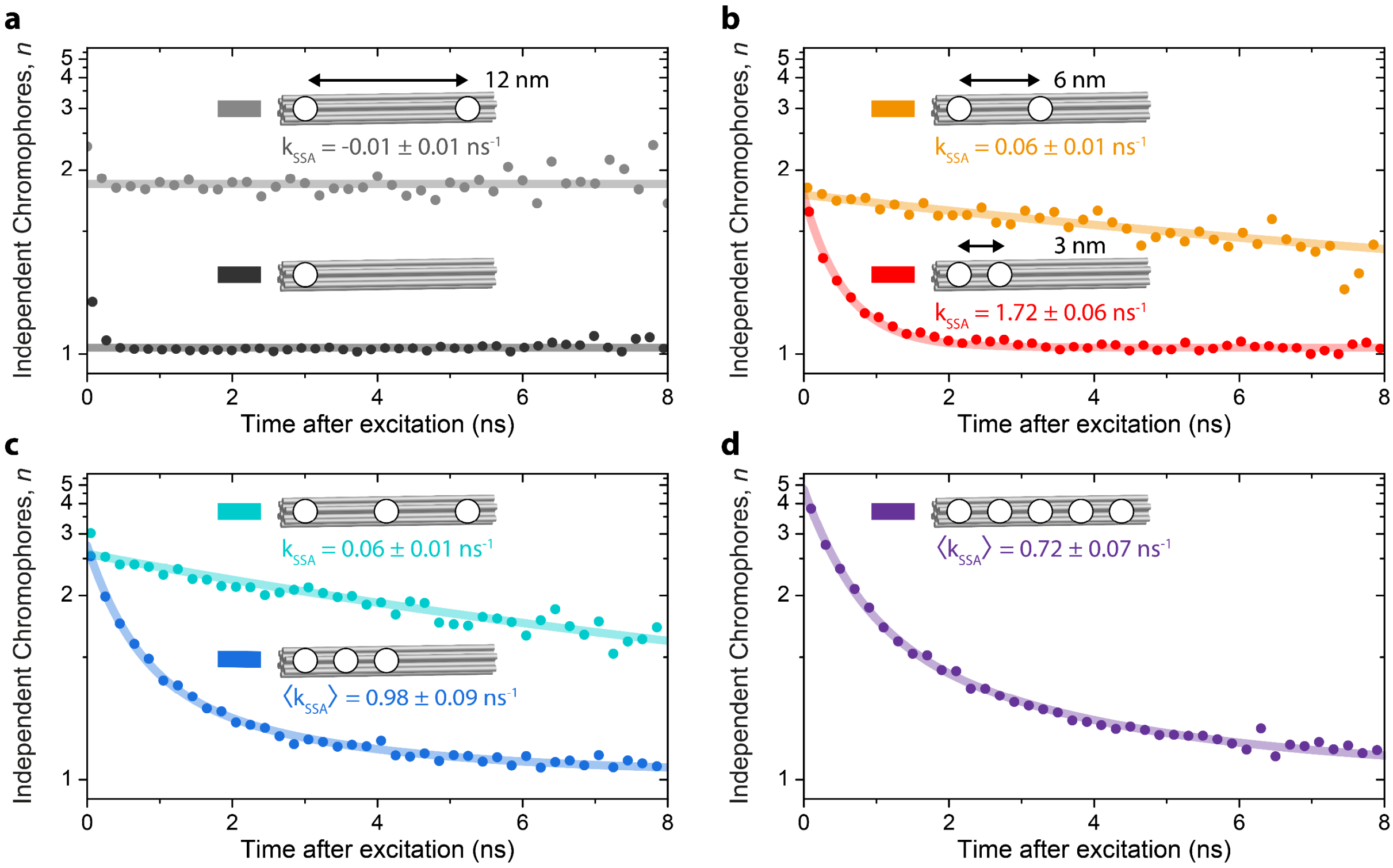

**Figure 3. Exciton annihilation in different well-defined mcNPs.** Evolution of the number of independently emitting chromophores, $n$, as a function of the time after excitation for seven different structures. The structures have different numbers of dyes attached with different spacings between them. The y-axes are reciprocal for better comparison between the graphs. (a) One dye (dark grey) and two dyes separated by 12 nm (light grey). (b) Two dyes separated by 6 nm (orange) and 3 nm (red). The curves superimposed are described by a monoexponential model of exciton annihilation (eq. 2). (c) Three dyes separated by 6 nm (cyan) and 3 nm (blue). The cyan curve is described by eq. 2, but the blue points follow a biexponential decay with an average rate $\langle k_{SSA} \rangle$ (eq. 3). (d) Five dyes separated by 3 nm. The curve is described by the biexponential decay of eq. 3. Between 54 and 98 single mcNPs were measured individually, and the photon statistics of each measurement accumulated to obtain each curve. All measurements were performed under oxygen removal and with a reducing and oxidizing system (ROXS) present to ensure photostabilization[27]. Each particle was measured for only 5 seconds so that photobleaching and spectral shifts were negligible.

Having established that we can recover the number of dyes in an mcNP with our method, we now apply this approach to different DNA origami structures to examine the dynamics of the SSA mechanism in detail. Figure 3 plots the number of independently emitting chromophores $n$ for each 200 ps time gate versus the corresponding microtime for seven different DNA origami structures. We start with the simplest model system with only one dye (dark grey dots in panel a). Except for the first two data points, these values are constant at $n$ = 1.02, which is expected because SSA cannot occur. This value is close to unity and only limited by the signal-to-background ratio (SBR) as discussed in Figure S5.[15] The fast decay in the first two data points originates from multiple excitations of the dye within the same laser pulse of ~80 ps width.[30] Now we introduce a second dye at a distance of ~12 nm (panel b, light grey dots), which should be large enough to prevent SSA between the excitons. Indeed, the data can be described with a constant $n$ of $1.85 \pm 0.01$, which is slightly below the expected value of two, most likely because of slightly different PL intensities of the two dye molecules at the different binding sites of the DNA origami structure. Crucially, again, no decay of $n$ is observed for this sample, implying a negligible exciton annihilation rate.

Next, we examine the more interesting cases, where we build structures with two dyes sufficiently close to each other such that SSA can occur. The red and orange dots in Figure 3b display the data measured on structures carrying two dyes at ~3 nm and ~6 nm spacing. $n$ starts out slightly below the expected value of two for both samples, and a decay during the first 2 ns down to $n = 1.02$ is observed for the 3 nm sample. These datasets are accurately described by a single-exponential model of the number of independently emitting chromophores,

$$n(t) = \{y_0 - [A \cdot \exp(-k_{\text{SSA}}\, t)]\}^{-1} \qquad \text{eq. (2)},$$

with the offset, $y_0$, amplitude, $A$, and the exciton annihilation rate, $k_{\text{SSA}}$, (see Methods for a derivation of eq. 2). The overall number of physical dyes present in the structure is then given by $n_{\text{dyes}} = (y_0 - A)^{-1}$. In Figure 3b, we extract $k_{\text{SSA}} = 1.72 \pm 0.06\ \text{ns}^{-1}$ for the two dyes separated by 3 nm and $k_{\text{SSA}} = 0.06 \pm 0.01\ \text{ns}^{-1}$ for the dyes separated by 6 nm, with $n_{\text{dyes}} = 1.8 \pm 0.03$ in both cases. As expected, $k_{\text{SSA}}$ drops significantly when doubling the distance between the two dyes, indicating that we are in the important regime where SSA is controlled by FRET and therefore by dye spacing. Subsequently, we placed three dyes separated by ~6 nm each (Figure 3c, cyan dots). Fitting with eq. 2 yields $k_{\text{SSA}} = 0.06 \pm 0.01\ \text{ns}^{-1}$ and $n_{\text{dyes}} =$ 2.7 $\pm$ 0.1, which is consistent because we expect no SSA between the left-most and right-most dyes, and the same SSA rate for the neighbouring dyes as in panel b.

Upon moving the three dyes closer to each other, now only separated by 3 nm (Figure 3c, blue dots), eq. 2 is no longer sufficient to describe the time evolution of $n$ since next-nearest neighbour interactions arise. We therefore used an analogous biexponential model of SSA, with a fast rate for neighbouring dyes and a slow rate for next-neighbouring dyes, which is a combination of direct annihilation and exciton hopping, to describe the blue dataset in panel c

$$n(t) = \{y_0 - [A_1 \exp(-k_{\mathrm{SSA,1}} \cdot t) + A_2 \exp(-k_{\mathrm{SSA,2}} \cdot t)]\}^{-1} \qquad \text{eq. (3).}$$

We derive from this dynamics an average amplitude-weighted SSA-rate $\langle k_{\mathrm{SSA}} \rangle = (A_1 k_{\mathrm{SSA,1}} + A_2 k_{\mathrm{SSA,2}})/(A_1 + A_2) = 0.98 \pm 0.09\ \mathrm{ns}^{-1}$ (see Supplementary Information for complete fitting results in Table S2) and an number of dyes, $n_{dyes} = (y_0 - (A_1 + A_2))^{-1} = 2.9\ \pm 0.1$. Finally, for the DNA origami structure bearing all five dyes (Figure 3d, violet dots), we extract $\langle k_{\mathrm{SSA}} \rangle = 0.72 \pm 0.07\ \mathrm{ns}^{-1}$ and $n_{dyes} = 4.7\ \pm 0.2$ by using eq. 3.

The crucial observation is that at long microtimes, $n$ decays to 1 for all samples with $k_{\mathrm{SSA}} > 0$. This is particularly intriguing for the five-dye sample, where we would anticipate the case in which two excitons remain on the left-most and right-most dyes. According to the experiment with two dyes placed 12 nm apart (panel a, light grey dots), no direct SSA should occur in this case. However, the fact that the five-dye sample still decreases down to only one emitting independent chromophore, rather than two, allows us to conclude that exciton hopping, i.e. exciton diffusion, occurs between the five dyes. This finding provides a motivation to study different mcNPs in which significant exciton diffusion arises.

**Exciton diffusion and annihilation in multichain aggregates of conjugated polymers.** To examine exciton diffusion in conjugated polymers in the mesoscopic size regime, aggregates of chains were grown with distinct electronic and structural properties. These structures are formed by two poly(*para*-phenylene-ethynylene-butadiynylene) (PPEB)-based conjugated polymers (Figure 4a). With a small variation of the alkyl side-chains, ordered aggregates with either H-type *interchromophoric* coupling (PPEB-1, lilac), or disordered aggregates with J-type *intrachromophoric* coupling (PPEB-2, brown) can be grown by solvent vapour annealing.[18] Samples were prepared as described in ref. [18], yielding individual small aggregates isolated in poly(methylmethacrylate) (PMMA) and measured on a confocal fluorescence microscope as described previously[31] and reviewed briefly in the Methods section. 631 single aggregates of PPEB-1, each comprising on average approximately 54 chains, and 705 aggregates of PPEB-2 (each ~9 chains, see Supplementary Information), were grown and measured individually. Following the above procedure, $n(t)$ was determined using psTRAB as shown in Figure 4b. We observe a clear decay of $n$ with time, signifying excited-state interactions primarily due to SSA. A substantial difference between the decay dynamics exists for the two aggregates. For the H-type aggregates, $n$ drops rapidly over the first 250 ps and then continues before levelling off at ~2,000 ps. The J-type aggregates show a smaller initial fast drop, followed by a slower linear decay before levelling off at a slightly higher value of $n$ at times >2,000 ps.

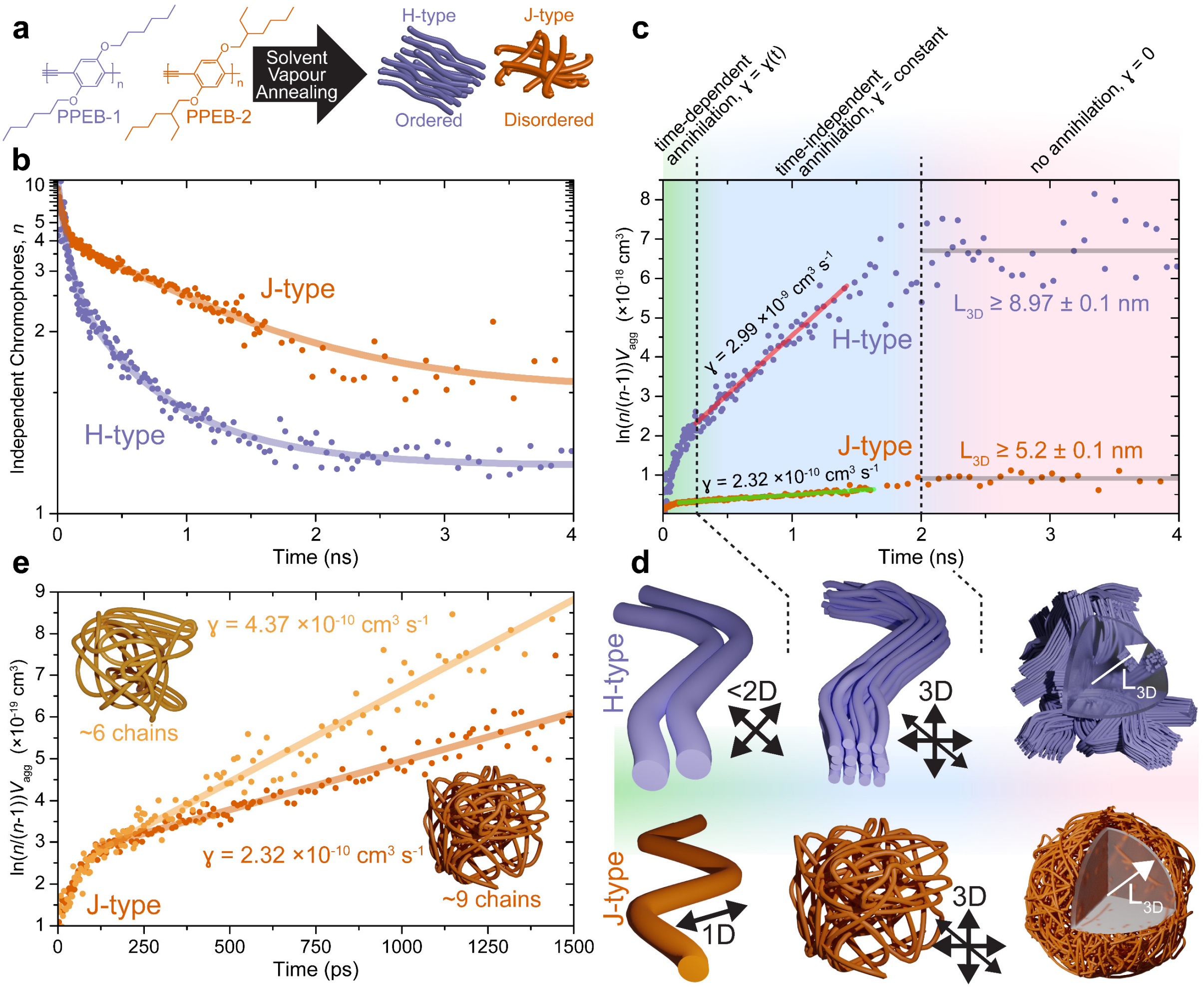

**Figure 4. Following exciton diffusion dynamics in isolated conjugated polymer aggregates.** (a) PPEB polymers that are grown into H-type (PPEB-1, lilac) or J-type (PPEB-2, brown) aggregates by solvent vapour annealing. (b) Temporal evolution of the number of independent chromophores in the aggregate, determined by psTRAB, for the H-type and J-type aggregates. A significant difference in the dynamics is observed, with lines being guides for the eye. (c) Plotting of the time-dependent quantity $\ln\left(\frac{n}{n-1}\right) \cdot V_{\mathrm{agg}}$ for the H- and J-type aggregates. The gradients of the curves correspond to the exciton annihilation rate, γ, of diffusion-controlled annihilation. Three regions of the dynamics are identified: at early time (0-250 ps), γ is time-dependent; at intermediate times (250-2,000 ps) γ is constant and tenfold higher in the H-compared to the J-type aggregate; and at late times (> 2,000 ps) γ is zero. These regions are interpreted in (d), indicating that early-time diffusion is one- or less-than two-dimensional and intermediate time diffusion is three-dimensional; at late times annihilation ceases because the exciton density is too low. The latter range provides a lower limit of the three-dimensional diffusion length, $L_{3\mathrm{D}}$. In (e) the nine-chain J-type aggregate shown in panel c (brown) is compared with a smaller six-chain J-type aggregate (orange). The gradient (γ) is a factor of two smaller in the larger aggregate, indicating that three-dimensional diffusivity is reduced when more chains are present. This reduction is consistent with reduced ordering of the chains as the aggregate size grows, and thus reduced interchain coupling. The six-chain dataset is made up of 260 individually measured single aggregates.

First, we note that, in analogy to the DNA origami model system in Figure 3d, the decay of $n$ with time constitutes a signature of exciton annihilation mediated by exciton diffusion. Because diffusion is now likely to dominate, however, the dynamics generally cannot be fitted with one fixed $k_{\mathrm{SSA}}$ rate. Instead, the annihilation is governed by a rate equation for a second-order reaction.[32] The clear difference between the H- and J-type aggregates indicates that the process of exciton diffusion is not the same in both of them. To examine this difference in a quantifiable manner, we plot the evolution with time of the quantity $\ln\left(\frac{n}{n-1}\right) \cdot V_{\mathrm{agg}}$ as shown in Figure 4c, where $V_{\mathrm{agg}}$ is the calculated aggregate volume (see Methods for a full description of this equation and Supplementary Information for how the volumes were obtained). This allows us to quantify and compare exciton diffusion, as data plotted in this manner allows the instantaneous rate of bimolecular exciton annihilation, $\gamma$, to be determined from the slope and compared against ensemble equivalents. A linear function signifies a constant, time-independent $\gamma$, whereas curvature implies that $\gamma$ has a time-dependence. Typically, in exciton annihilation measurements, the underlying excited-state decay has to be accounted for,[33] complicating analysis in extracting diffusion relevant properties. The advantage of psTRAB is that we directly obtain a measure of the exciton diffusion and are thus uniquely sensitive to weak and slow diffusion. This contrasts with conventional ensemble measurements of the non-exponential decay in PL intensity, which require high exciton densities to see an appreciable effect of annihilation. With the data plotted as $\ln\left(\frac{n}{n-1}\right) \cdot V_{\mathrm{agg}}$ as in Figure 4c, for both H- and J-type aggregates three regions are identified. At early times (<250 ps) non-linear behaviour is observed, indicating that $\gamma$ is time dependent. Exciton diffusion is therefore one- or less-than two-dimensional.[34] At times 250-2,000 ps, both aggregate types show linear behaviour, thus $\gamma$ is time-independent and the diffusion three-dimensional,[35, 36] with values of $\gamma$ found to be in the range of $10^{-9}$ to $10^{-10}$ $\mathrm{cm^3\ s^{-1}}$, in good agreement with typical conjugated polymers.[29, 36-38]

Finally, at times >2,000 ps, $\gamma = 0$, i.e. annihilation has ceased as the exciton density is too low to support continued interactions.

The psTRAB results also allow insight into the nanoscale organisation of material in the aggregates, as sketched in Figure 4d. At early times, the time-dependent $\gamma$ indicates that exciton motion is one- or less-than two-dimensional, most likely in the dispersive regime, and is therefore consistent with ensemble observations of annihilation on the timescale of a few picoseconds.[39] In the context of the H-type aggregate, this motion will be along the chains and across the interchain $\pi$-stack. This conclusion is in agreement with a high degree of chain alignment, evidenced by the PL intensity modulation depths determined when rotating the polarization of the exciting laser.[18] The J-type aggregate also shows time-dependent annihilation at early times. Here, however, simple one-dimensional motion will be favoured since strong intrachain coupling is dominant as evidenced by the J-type emission characteristics.[18] At later times, the time-independence of $\gamma$ indicates that exciton motion is three-dimensional in both aggregate types. $\gamma$ is an order of magnitude lower in this time region for the J than for the H-type. This difference relates to the nature of chromophoric coupling and disorder in the aggregates. In H-type aggregates, chains with the smallest degree of disorder will show the strongest interchain electronic coupling, facilitating efficient three-dimensional diffusion. In J-type aggregates, in contrast, which do not show a high polarization anisotropy[18], chains are relatively disordered. Poor chain alignment will lead to weak interchain electronic coupling and a lower value of $\gamma$. Exciton diffusion is then limited by the random chain alignment that excitons encounter when diffusing. The impact of chain disorder on exciton diffusion can also be examined by comparing the psTRAB of the 9-chain J-type aggregate with a smaller one that comprises of ~6 chains shown in Figure 4e. In the region where $\gamma$ is time-independent and three-dimensional diffusion dominates, $\gamma$ is almost a factor of two higher in the 6-chain aggregate, indicating increased order in the smaller aggregate which facilitates effective interchain site-to-site hopping.

Finally, at late times where $\gamma \rightarrow 0$, we enter the regime where the exciton density is too low to support continued annihilation. These conditions can be used to obtain a lower limit on the exciton diffusion length, $L_{3D}$. The rationale for this approach is simple: we know the volume of the aggregate and the number of independent chromophores that the aggregate can support when we can no longer measure annihilation occurring, i.e. when excitons no longer interact with each other. Division yields the volume that a single independent chromophore occupies, equivalent to the volume explored by an exciton. If diffusion is presumed to arise in a spherical volume in three dimensions, a diffusion length, $L_{3\mathrm{D}}$, can be determined. The value will be a lower limit as the length is technically defined as the distance excitons diffuse in their lifetime rather than once the exciton density is too low to support continued interactions, but the difference between these two definitions will be small at these late times. We find lower limits of $L_{3\mathrm{D}} \approx 9$ nm for the H-type aggregate and $L_{3\mathrm{D}} \approx 5.2$ nm for the J-type aggregate, consistent with typical literature values for conjugated polymers.[23, 33, 36, 40] The unique advantage of our chromophore-counting method is that the calculation of these values contains no presumptions other than the mass density of the aggregate. $L_{3\mathrm{D}}$ is derived from simple observables and is only possible because we consider single objects at the discretised level of excitons and the resulting photon correlation.

**Conclusions**

Knowledge of the nanoscale organisation of a material, the electronic coupling between chromophores and energy transfer pathways is important in a wide variety of systems. In this work we have introduced a new method to quantify exciton-exciton annihilation and exciton diffusion in multichromophoric mesoscopic objects. This is achieved by resolving the fluorescence photon statistics on a picosecond timescale. Using deterministic DNA origami structures, we position dyes at specific distances from each other and obtain direct measurements of the rate of annihilation between two excitons and the true number of dyes. This accuracy is a direct consequence of utilising two-detector coincidences that are sensitive to two-photon emission events. Our method can measure the annihilation rate $\gamma$ in well-defined structures and directly yields the number of physical dyes present in each sample. We stress that such chromophore counting is not possible with standard time-integrated photon-correlation measurements. The technique can be expanded to look at nanoparticles grown from multiple single conjugated-polymer chains. In these polymer aggregates, SSA is governed mainly by exciton diffusion instead of fixed distance FRET-based annihilation between chromophores. In addition, the method offers facile differentiation between J- and H-type aggregates, determining valuable material properties such as the exciton diffusion length, the dimensionality of diffusion and the degree of nanoscale disorder in the aggregate. The technique therefore offers valuable new opportunities to explore the nanoscale organisation and excitonic coupling of chromophores in light emitting materials with unprecedented detail.

## Methods

**Photon correlation, data analysis and derivation of eq. 2:** The psTRAB is computed from raw time-stamped TCSPC data using MATLAB. The scripts developed operate similarly to conventional calculations of cross-correlations.[41] The following parameters are stored for each photon event: (i) the “macrotime” at which the photon arrived, i.e. the integer multiple of the corresponding excitation laser repetition period $T$; (ii) the “microtime”, $t$, which corresponds to the time the photon was detected after the excitation pulse excited the NP; and (iii) the detection channel, i.e. the photon counter $A$ or $B$. The events are cross-correlated with respect to their macrotimes, after which the microtimes are evaluated as follows: (i) we store the shorter microtime, $t$, of each correlation event (e.g. the microtime of channel $A$) and neglect the longer microtime, $t + \Delta t$. (ii) For selected microtime intervals, histograms of correlation events are constructed as a function of the macrotime delay between the channels. Finally, the scripts sum over multiple measurements of individual aggregates to produce an overall psTRAB result. As detailed in the supplementary information, we rationalize the number of correlation events, $N_\mathrm{c}(t, t+\Delta t)$, for a given delay time $\Delta t < T - t$ between two photon events arising from the same excitation pulse, as follows:

$$N_\mathrm{c}(t, t+\Delta t) = N_\mathrm{exc} \cdot P(t) \cdot P'(t+\Delta t) \qquad \text{eq. (4)}$$

Here, $N_\mathrm{exc}$ is the total number of observed laser excitation pulses, $P(t)$ is the probability of detecting the first photon at microtime $t$ and $P'(t+\Delta t)$ is the probability of detecting the second photon at microtime $t+\Delta t < T$. In case the exciton annihilation is determined by a single exponential decay rate $k_\mathrm{SSA}$, these probabilities are calculated as

$$P(t) = n_\mathrm{dyes} p_0 e^{-(k_\mathrm{r}+k_\mathrm{nr}+k_\mathrm{ET})t} \qquad \text{eq. (5)}$$

$$P'(t+\Delta t) = \left(n_\mathrm{dyes} - 1\right) p_0 e^{-(k_\mathrm{r}+k_\mathrm{nr}+k_\mathrm{ET})t} e^{-(k_\mathrm{r}+k_\mathrm{nr})\Delta t}, \qquad \text{eq. (6)}$$

where $n_\mathrm{dyes}$ is the number of chromophores, $p_0$ summarizes the probability of the chromophore being excited by the laser pulse and the probability of detecting the emitted photon, $k_\mathrm{r}$ and $k_\mathrm{nr}$ are the radiative and non-radiative decay rates and $k_\mathrm{ET} = k_\mathrm{SSA}/2$ is the energy-transfer rate between two excited chromophores. Note that in general $P'(t) \neq P(t)$ since the exciton emitting the first photon at time $t$ can reside on any one of the $n_\mathrm{dyes}$ chromophores, while the exciton emitting the second photon resides on one of the $(n_\mathrm{dyes} - 1)$ remaining chromophores. At microtime delays $0 < \Delta t < T - t$, the number of excitons does not decay any further through energy transfer, since only a single exciton is left. The number of correlation events $N_\mathrm{l}(t, t+\Delta t)$, where the second photon is detected at non-zero macrotime delays and thus arises due to a separate laser excitation event, is instead calculated from

$$N_\mathrm{l}(t, t+\Delta t) = N_\mathrm{exc} \cdot P''(t) \cdot P''(t+\Delta t) \qquad \text{eq. (7),}$$

where

$$P''(t) = n_{\text{dyes}} p_0 e^{-(k_\text{r}+k_\text{nr})t} \quad \text{eq. (8)}$$

is independent of energy transfer, since only single excitons are present after each laser excitation. The ratio $N_\text{c}/N_\text{l}$ of central to lateral correlation events is thus directly connected to the number of chromophores in the mcNP and the time dynamics of the annihilation process as

$$\frac{N_\text{c}}{N_\text{l}} = \frac{n_{\text{dyes}}(n_{\text{dyes}}-1)}{n_{\text{dyes}}{}^2} e^{-k_{\text{SSA}}t} = \frac{n_{\text{dyes}}-1}{n_{\text{dyes}}} e^{-k_{\text{SSA}}t} \quad \text{eq. (9)}$$

Note that the result is independent of $\Delta t$ and it can also be calculated from the time-integrated number of correlations

$$N_\text{c}(t) = \int_0^{T-t} N_\text{c}(t, t+\Delta t) d(\Delta\text{t}), \quad N_\text{l}(t) = \int_0^{T-t} N_\text{l}(t, t+\Delta t) d(\Delta\text{t}) \quad \text{eq. (10)}$$

which significantly reduces the noise associated with experimental event data.

Comparing the derived expression for $N_\text{c}/N_\text{l}$ with eq. 1 defining the number of independent chromophores $n$, we obtain

$$n(t) = \left(1 - \frac{n_{\text{dyes}}-1}{n_{\text{dyes}}} \exp(-k_{\text{SSA}}t)\right)^{-1} \quad \text{eq. (11).}$$

Eq. 11 corresponds to eq. 2 with $y_0 = 1$ and $A = 1 - n_{\text{dyes}}^{-1}$. A quantum-statistical description of photon correlations in an n-chromophore system, using a master equation approach, is given in the Supplementary Information. Note that the assumption of any specific decay law for singlet-singlet annihilation such as an exponential decay according to $e^{-k_{\text{SSA}}t}$ is not strictly necessary. To that end, psTRAB $N_\text{c}/N_\text{l}$ can be used to directly measure the decay law associated with exciton-exciton interactions, which is connected to the mean first passage time of the random walk performed by the excitons. The technique can obviously be extended to higher-order photon correlations, using more than one beam splitter in the Hanbury Brown and Twiss setup, to determine the functional difference between two-exciton interactions and higher-order contributions.

**DNA Origami Microscopy:** A custom-made confocal microscope based on an Olympus IX-71 inverted microscope was used. Multichromophoric DNA-origami structures were excited by a pulsed laser (636 nm, ~80 ps full-width half-maximum, 80 MHz, LDH-D-C-640; PicoQuant GmbH) operated at 40 MHz repetition rate. Circularly polarized light was obtained by a linear polarizer (LPVISE100-A, Thorlabs GmbH) and a quarter-wave plate (AQWP05M-600, Thorlabs GmbH). The light was focused onto the sample by an oil-immersion objective (UPLSAPO100XO, NA 1.40, Olympus Deutschland GmbH). The sample was moved by a piezo stage (P-517.3CD, Physik Instrumente (PI) GmbH & Co. KG) controlled by a piezo controller (E-727.3CDA, Physik Instrumente (PI) GmbH & Co. KG). The emission was separated from the excitation beam by a dichroic beam splitter (zt532/640rpc, Chroma) and focused onto a 50 μm pinhole (Thorlabs GmbH). The emission light was separated from scattered excitation light by a 647 nm long-pass filter (RazorEdge LP 647, Semrock) and split into two detection channels by a non-polarizing 50:50 beam splitter (CCM1-BS013/M, Thorlabs GmbH). In each detection channel, afterglow of the avalanche photodiode was blocked by a 750 nm short-pass filter (FES0750, Thorlabs GmbH). Emission was focused onto avalanche photodiodes (SPCM-AQRH-14-TR, Excelitas Technologies GmbH & Co. KG) and signals were registered by a multichannel picosecond event timer (HydraHarp 400, PicoQuant GmbH). The setup was controlled by a commercial software package (SymPhoTime64, Picoquant GmbH).

**PPEB Aggregate Microscopy:** Single polymer aggregates were measured on a custom-designed confocal microscope as described previously.[31] For excitation, the frequency-doubled output of a Ti:Sapphire oscillator (~100 fs, 80 MHz, 810 & 880 nm) (Chameleon, Coherent) was used, centred at 405 nm for PPEB-1 and 440 nm for PPEB-2. Femtosecond excitation was required to ensure that double excitation of the aggregates did not occur, because the excited state lifetime for the J-type coupled PPEB-2 aggregates is significantly shorter than for the DNA-origami dyes,[18] preventing the use of conventional picosecond laser diodes. The laser was spatially expanded, spectrally cleaned and coupled into the microscope base (IX71, Olympus K.K., Japan), where it filled the backplane of a 60× 1.35 NA objective (UPLSAPO60XO, Olympus K.K., Japan). The sample was placed on a piezo stage (P-527.3CL, Physik Instrumente GmbH, Germany) which was scanned to generate microscope images and locate individual aggregates. The PL was detected using two single-photon detectors (PD-25-CTE, Micro Photon Devices S.r.l., Italy) connected to a multichannel picosecond event timer (HydraHarp 400, PicoQuant GmbH, Germany) allowing time-correlated single-photon counting and cross-correlations to be performed. The piezo stage and photon counting hardware were controlled using a customized code in LabVIEW (National Instruments).

**Exciton Diffusion in PPEB Aggregates:** Bulk exciton-exciton annihilation by SSA is conventionally described by a simple second-order reaction equation, $\frac{d}{dt}\rho_{\mathrm{exc}} = -\gamma(t)\rho_{\mathrm{exc}}^2$, where $\rho_{\mathrm{exc}}$ is the exciton density and $\gamma(t)$ is the diffusion-controlled annihilation rate. In the context of our psTRAB method, differentiation of equation 11 ultimately leads to

$$\frac{d}{dt}n = -k_{\mathrm{SSA}} \cdot n(n-1) \qquad \text{eq. (12).}$$

for the number of independent chromophores. This function is the correct form of the second-order reaction equation in cases where the number of reactants is low, since the reaction rate of change is proportional to the number of pairs that can be chosen. The psTRAB measurements thus resolves SSA on the single-nanoparticle level in a form that can be thought of qualitatively as tracking the *mutual annihilation of independent chromophores* by bimolecular interaction. From equation 12, we derive the following linear form governing the exciton annihilation rate $\gamma = k_{\mathrm{SSA}} \cdot V_{\mathrm{agg}}$, where $V_{\mathrm{agg}}$ is the aggregate volume:

$$-V_{\mathrm{agg}} \cdot \ln\left(\frac{n-1}{n}\right) = \gamma \cdot t - V_{\mathrm{agg}} \cdot \ln\left(\frac{n_0-1}{n_0}\right) \qquad \text{eq. (13).}$$

See the Supplementary Information for details on how $V_{\mathrm{agg}}$ is obtained by simply invoking knowledge of the mass and mass density of the polymer chain and the number of chains in the aggregate. Thus, plotting $\ln\left(\frac{n}{n-1}\right) \cdot V_{\mathrm{agg}}$ as a function of $t$ as in Figure 4c,e allows $\gamma$ to be determined from the gradient by straight-line fitting.

## Acknowledgements

TE and FH thank the Deutsche Forschungsgemeinschaft (German Research Foundation) for funding through Collaborative Grant No. 319559986. FH thanks the Deutsche Forschungsgemeinschaft for funding through Project-ID 314695032-SFB 1277. PT, FS and TS thank the European Union's Horizon 2020 research and innovation programme under grant agreement No 737089 (Chipscope) and the DFG (excellence cluster e-conversion) for financial support. We thank Dr. Florian Selbach for TEM imaging.

## Author Contributions

GJH, FS, FH and JV devised the psTRAB methodology and developed the technique. TS, JV and PT designed the DNA origami structures. TS prepared, measured & analysed the DNA origami structures and data. D.L. and S.H. designed and synthesized the conjugated polymers. GJH, FS and TE measured & analysed the PPEB aggregate data. SB developed the analytic and quantum-statistical treatment of psTRAB. All authors contributed to manuscript writing.

## Competing Interests statement

The authors declare no competing financial interests.

## References

1. Hollars, C. W., Lane, S. M. & Huser, T. Controlled non-classical photon emission from single conjugated polymer molecules. *Chem. Phys. Lett.* **370**, 393-398 (2003).

2. Stangl, T., Wilhelm, P., Remmerssen, K., Höger, S., Vogelsang, J. & Lupton, J. M. Mesoscopic quantum emitters from deterministic aggregates of conjugated polymers. *Proc. Natl. Acad. Sci. U.S.A.* **112**, E5560-E5566 (2015).

3. Utzat, H., Shulenberger, K. E., Achorn, O. B., Nasilowski, M., Sinclair, T. S. & Bawendi, M. G. Probing Linewidths and Biexciton Quantum Yields of Single Cesium Lead Halide Nanocrystals in Solution. *Nano Lett.* **17**, 6838-6846 (2017).

4. Nair, G., Zhao, J. & Bawendi, M. G. Biexciton Quantum Yield of Single Semiconductor Nanocrystals from Photon Statistics. *Nano Lett.* **11**, 1136-1140 (2011).

5. Sun, C.*, et al.* Biexciton Binding of Dirac fermions Confined in Colloidal Graphene Quantum Dots. *Nano Lett.* **15**, 5472-5476 (2015).

6. Mangum, B. D., Ghosh, Y., Hollingsworth, J. A. & Htoon, H. Disentangling the effects of clustering and multi-exciton emission in second-order photon correlation experiments. *Opt. Express* **21**, 7419-7426 (2013).

7. Trofymchuk, K.*, et al.* Giant light-harvesting nanoantenna for single-molecule detection in ambient light. *Nat. Photon.* **11**, 657-663 (2017).

8. Wientjes, E., Renger, J., Curto, A. G., Cogdell, R. & van Hulst, N. F. Strong antenna-enhanced fluorescence of a single light-harvesting complex shows photon antibunching. *Nat. Commun.* **5**, 4236 (2014).

9. Tinnefeld, P.*, et al.* Antibunching in the emission of a single tetrachromophoric dendritic system. *J. Am. Chem. Soc.* **124**, 14310-14311 (2002).

10. Fisher, B., Caruge, J. M., Zehnder, D. & Bawendi, M. Room-temperature ordered photon emission from multiexciton states in single CdSe core-shell nanocrystals. *Phys. Rev. Lett.* **94**, 087403 (2005).

11. Steiner, F., Vogelsang, J. & Lupton, J. M. Singlet-Triplet Annihilation Limits Exciton Yield in Poly(3-Hexylthiophene). *Phys. Rev. Lett.* **112**, 137402 (2014).

12. Schröder, T., Scheible, M. B., Steiner, F., Vogelsang, J. & Tinnefeld, P. Interchromophoric Interactions Determine the Maximum Brightness Density in DNA Origami Structures. *Nano Lett.* **19**, 1275-1281 (2019).

13. Hedley, G. J., Ruseckas, A. & Samuel, I. D. W. Light Harvesting for Organic Photovoltaics. *Chem. Rev.* **117**, 796-837 (2017).

14. Tinnefeld, P., *et al.* Higher-excited-state photophysical pathways in multichromophoric systems revealed by single-molecule fluorescence spectroscopy. *ChemPhysChem* **5**, 1786-1790 (2004).

15. Weston, K. D., Dyck, M., Tinnefeld, P., Müller, C., Herten, D. P. & Sauer, M. Measuring the number of independent emitters in single-molecule fluorescence images and trajectories using coincident photons. *Anal. Chem.* **74**, 5342-5349 (2002).

16. Ta, H., Wolfrum, J. & Herten, D. P. An Extended Scheme for Counting Fluorescent Molecules by Photon-Antibunching. *Laser Phys.* **20**, 119-124 (2010).

17. Grußmayer, K. S. & Herten, D. P. Time-resolved molecule counting by photon statistics across the visible spectrum. *Phys. Chem. Chem. Phys.* **19**, 8962-8969 (2017).

18. Eder, T., *et al.* Switching between H- and J-type electronic coupling in single conjugated polymer aggregates. *Nat. Commun.* **8**, 1641 (2017).

19. Hofkens, J., *et al.* Revealing competitive Forster-type resonance energy-transfer pathways in single bichromophoric molecules. *Proc. Natl. Acad. Sci. U.S.A.* **100**, 13146-13151 (2003).

20. Hübner, C. G., Zumofen, G., Renn, A., Herrmann, A., Müllen, K. & Basché, T. Photon antibunching and collective effects in the fluorescence of single bichromophoric molecules. *Phys. Rev. Lett.* **91**, 093903 (2003).

21. Fuckel, B., Hinze, G., Nolde, F., Müllen, K. & Basché, T. Quantification of the Singlet-Singlet Annihilation Times of Individual Bichromophoric Molecules by Photon Coincidence Measurements. *J. Phys. Chem. A* **114**, 7671-7676 (2010).

22. Stangl, T., *et al.* Temporal Switching of Homo-FRET Pathways in Single-Chromophore Dimer Models of pi-Conjugated Polymers. *J. Am. Chem. Soc.* **135**, 78-81 (2013).

23. Lin, J. D. A., *et al.* Systematic study of exciton diffusion length in organic semiconductors by six experimental methods. *Mater. Horiz.* **1**, 280-285 (2014).

24. Daniel, C., *et al.* Exciton bimolecular annihilation dynamics in supramolecular nanostructures of conjugated oligomers. *Phys. Rev. B* **68**, 235212 (2003).

25. Derr, N. D., Goodman, B. S., Jungmann, R., Leschziner, A. E., Shih, W. M. & Reck-Peterson, S. L. Tug-of-War in Motor Protein Ensembles Revealed with a Programmable DNA Origami Scaffold. *Science* **338**, 662-665 (2012).

26. Schmied, J. J., *et al.* DNA origami-based standards for quantitative fluorescence microscopy. *Nat. Protoc.* **9**, 1367-1391 (2014).

27. Vogelsang, J., *et al.* A reducing and oxidizing system minimizes photobleaching and blinking of fluorescent dyes. *Angew. Chem. Int. Edit.* **47**, 5465-5469 (2008).

28. Grußmayer, K., Yserentant, K. & Herten, D. P. Photons in - numbers out: perspectives in quantitative fluorescence microscopy for in situ protein counting. *Methods Appl. Fluoresc.* **7**, 012003 (2019).

29. Gulbinas, V., Mineviciute, I., Hertel, D., Wellander, R., Yartsev, A. & Sundstrom, V. Exciton diffusion and relaxation in methyl-substituted polyparaphenylene polymer films. *J. Chem. Phys.* **127**, 144907 (2007).

30. Schedlbauer, J., *et al.* Ultrafast Single-Molecule Fluorescence Measured by Femtosecond Double-Pulse Excitation Photon Antibunching. *Nano Lett.* **20**, 1074-1079 (2020).

31. Hedley, G. J., Steiner, F., Vogelsang, J. & Lupton, J. M. Fluctuations in the Emission Polarization and Spectrum in Single Chains of a Common Conjugated Polymer for Organic Photovoltaics. *Small* **14**, 1804312 (2018).

32. Alemany, P. A., Zanette, D. H. & Wio, H. S. Time-Dependent Reactivity for Diffusion-Controlled Annihilation and Coagulation in 2 Dimensions. *Phys. Rev. E* **50**, 3646-3655 (1994).

33. Mikhnenko, O. V., Blom, P. W. M. & Thuc-Quyen, N. Exciton diffusion in organic semiconductors. *Energy Environ. Sci.* **8**, 1867-1888 (2015).

34. Gulbinas, V., Chachisvilis, M., Valkunas, L. & Sundstrom, V. Excited state dynamics of phthalocyanine films. *J. Phys. Chem.* **100**, 2213-2219 (1996).

35. Shaw, P. E., Ruseckas, A., Peet, J., Bazan, G. C. & Samuel, I. D. W. Exciton-Exciton Annihilation in Mixed-Phase Polyfluorene Films. *Adv. Funct. Mater.* **20**, 155-161 (2010).

36. Shaw, P. E., Ruseckas, A. & Samuel, I. D. W. Exciton diffusion measurements in poly(3-hexylthiophene). *Adv. Mater.* **20**, 3516-3520 (2008).

37. King, S. M., Dai, D., Rothe, C. & Monkman, A. P. Exciton annihilation in a polyfluorene: Low threshold for singlet-singlet annihilation and the absence of singlet-triplet annihilation. *Phys. Rev. B* **76**, 085204 (2007).

38. Masri, Z., *et al.* Molecular Weight Dependence of Exciton Diffusion in Poly(3-hexylthiophene). *Adv. Energy Mater.* **3**, 1445-1453 (2013).

39. Maniloff, E. S., Klimov, V. I. & McBranch, D. W. Intensity-dependent relaxation dynamics and the nature of the excited-state species in solid-state conducting polymers. *Phys. Rev. B* **56**, 1876-1881 (1997).

40. Haugeneder, A., *et al.* Exciton diffusion and dissociation in conjugated polymer fullerene blends and heterostructures. *Phys. Rev. B* **59**, 15346-15351 (1999).

41. Gamari, B. D., *et al.* Inexpensive electronics and software for photon statistics and correlation spectroscopy. *Am. J. Phys.* **82**, 712-722 (2014).

Supplementary Information for

# Picosecond time-resolved antibunching measures nanoscale exciton motion, annihilation, and true number of chromophores

*Gordon J. Hedley[1,†,*], Tim Schröder[2,†], Florian Steiner[2], Theresa Eder[3], Felix Hofmann[3], Sebastian Bange[3], Dirk Laux,[4] Sigurd Höger,[4] Philip Tinnefeld[2], John M. Lupton[3] and Jan Vogelsang[3,*]*

[1]School of Chemistry, University of Glasgow, University Avenue, Glasgow, G12 8QQ, United Kingdom

[2]Department Chemie and Center for NanoScience (CeNS), Ludwig-Maximilians-Universität München, Butenandtstraße 5-13 Haus E, 81377 München, Germany

[3]Institut für Experimentelle und Angewandte Physik, Universität Regensburg, Universitätsstraße 31, 93053 Regensburg, Germany

[4]Kekulé-Institut für Organische Chemie und Biochemie, Universität Bonn, Gerhard-Domagk-Strasse 1, 53121 Bonn, Germany

## 1. Details of DNA origami structures and sample preparation

The DNA origami structure[1] was modified using caDNAno (version 0.2.2, design schematics in Fig. S1). The scaffold is an 8064 nucleotide long ssDNA extracted from M13mp18 bacteriophages. All staple strands were purchased from Eurofins Genomics GmbH as well as the dye labeled oligonucleotides (see at the end of Supplementary Information). The ATTO 542 modified oligonucleotides for external labeling were purchased from biomers.net. Scaffold and oligonucleotides were mixed according to table S1 for origami folding. The folding buffer (FB) is a Tris-EDTA buffer (1x TE, 10 mM Tris-HCl, 1 mM EDTA•$Na_2$) with 20 mM $MgCl_2$ and 5 mM NaCl. In the annealing process, the mixture was heated and slowly cooled down with a nonlinear thermal ramp over 16 hours according to reference.[2] After annealing, the excess staples were removed with polyethylene glycol (PEG) precipitation. The samples were mixed with an equal volume of PEG precipitation buffer (1× TAE, 15 % (w/v) PEG-8000, 500 mM NaCl, 12 mM $MgCl_2$) and centrifuged at 16 krcf (thousand relative centrifugal force, i.e. 1000 g) for 30 min at 4 °C. After removing the supernatant, the pellet was suspended in 1× FB. Afterwards, the DNA origami was externally labeled with ATTO 542 modified oligonucleotides. A threefold excess with respect to the extended staples was used and incubated for 20 min in a wet chamber at room temperature. The DNA origami structures were purified via gel electrophoresis. A 1.5 % agarose gel containing a Tris base, acetic acid and EDTA buffer (0.5× TAE, 20 mM Tris-HCl, 10 mM acetic acid, 0.5 mM EDTA) and 12 mM $MgCl_2$ was used at 60 V for 2 hours in a gel box cooled in an ice-water bath. The gel was not stained to avoid staining reagent-dye interactions. On a blue illuminated table DNA origami structures could be seen due to the numerous ATTO 542 dyes. DNA origami structures were recovered from the target band. The samples were stored at -26 °C until further use.

### Folding Table

Final concentrations for DNA origami folding are given in Table S1. The meaning of the reagents is described below:

**Table S1:** Folding reagents with final concentrations.

| **Reagent** | **Final concentration / nM** |
|---|---|
| scaffold | 25 |
| core staples | 225 |
| biotin staples | 250 |
| extended staples | 225 |
| dye and refill staples | 225 |

**scaffold:** Single-stranded viral 8064 nt ssDNA from M13mp18.

**core staples:** Contains every unmodified staples of the rectangular DNA origami. The wildtype structure is given in reference[1].

**biotin staples:** Four biotin modified staples. Modifications are placed at the 3' end.

**extended staples:** 13 staples extended at the 3' end for external labeling. The extended sequence is: 5' TTTTCCTCTACCACCTACATCAC 3'. Sequence for the ATTO542 oligonucleotides: 5' GTGATGTAGGTGGTAGAGGA-ATTO542 3' **dye and refill staples:** Oligonucleotides labeled with ATTO 647N at the 5' end. For structures which do not need all five dyes the respective oligonucleotides were substituted with unmodified oligonucleotides.

### Confirming successful DNA origami structure assembly with TEM imaging

Successful assembly of the modified DNA origami structure was confirmed with TEM imaging. The origami structures were imaged on Ar-plasma cleaned TEM grids (Formvar/carbon, 400 mesh, Cu, TedPella, Inc., USA). The DNA origami structures were stained with a 2 % uranyl formate solution. The imaging was performed on a JEM-1100 microscope (JEOL GmbH, Japan) with an acceleration voltage of 80 kV.

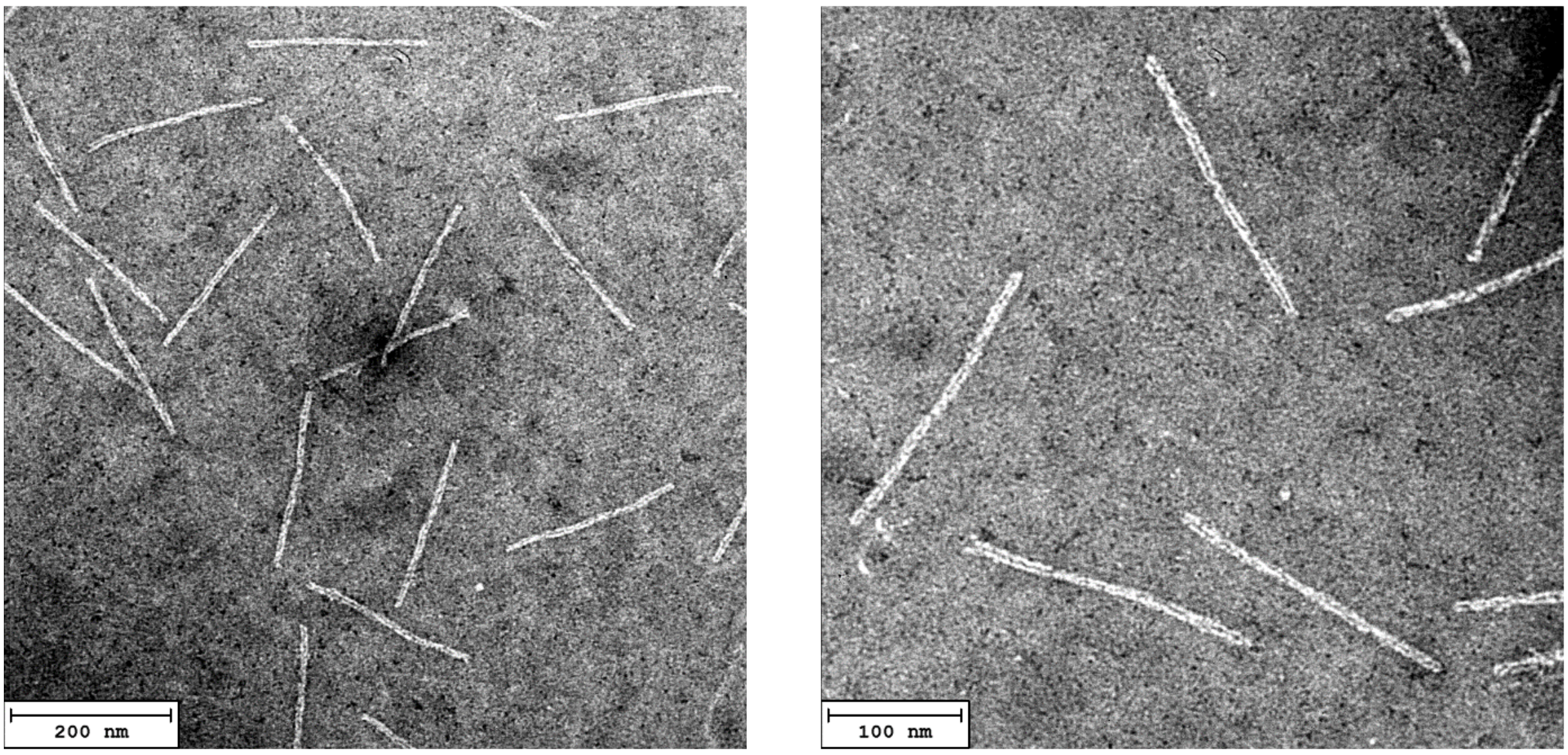

***Figure S1.*** *Two representative negative stain images of the modified DNA origami structure with different magnifications. The expected length of the DNA origami structure is 225 nm. The TEM images show successful assembly of the DNA origami structures. No nicks were found at the position which was modified for our needs and is located at the center of the structure (orange and blue staples in figure S2).*

### Surface preparation and immobilization

Samples were measured in LabTek™ chamber slides (Thermo Fisher Scientific Inc.) which were cleaned twice for 20 minutes with 0.1 M hydrofluoric acid (AppliChem GmbH) and washed afterwards three times with ultrapure water. For sample immobilization the glass surface was coated with biotin labeled bovine serum albumin (BSA) (1 mg/mL Sigma-Aldrich Chemie GmbH) and NeutrAvidin (1 mg/mL, Sigma-Aldrich Chemie GmbH). The DNA origami structures (60 pM in 1x FB) were immobilized by the biotin-NeutrAvidin binding.

### Sample preparation for single-molecule measurements

Traces were recorded from immobilized DNA origami structures. Photo blinking and photo bleaching was reduced with an oxidizing and reducing buffer system (1× TAE, 12 mM $MgCl_2$, 2 mM Trolox/Troloxquinone, 1 % (w/v) D-(+)-Glycose) [3] in combination with an oxygen scavenging system [4] (1 mg $mL^{-1}$ glucose oxidase, 0.4 % (v/v) catalase (50 µg $mL^{-1}$), 30 % glycerol, 12.5 mM KCl in 50 mM TRIS). The oxygen scavenging system was added to the oxidation and reducing buffer at a concentration of 10 % (v/v) in the LabTek$^{TM}$ system. DNA origami structures were excited with a laser power adjusted to 1 µW for all samples, except for the 5-dye sample which was measured with 0.5 µW to reduce spectral shifts and photo bleaching. The laser repetition rate was 40 MHz.

Traces for psTRAB data processing were recorded for 10 seconds. Origami structures with five dyes were recorded for 5 seconds.

### DNA origami structure

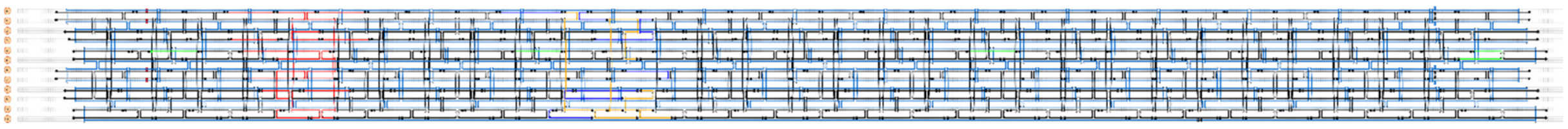

***Figure S2.*** *caDNAno sketch of the DNA origami structure used. Modified staples are colored. Green staples are biotin labeled. Red staples are extended at the 3' (5' TTTTCCTCTACCACCTACATCAC) end for external labeling with ATTO 542 modified oligonucleotides. Orange staples are labeled at the 5' end with ATTO 647N. Blue staples are next to the ATTO 647N labeled staples to stabilize the structure.*

### psTRAB data processing

Every trace was evaluated and only traces with constant fluorescence signal and without photobleaching were used for further data processing. This is necessary because singlet-singlet-annihilation leads to highly excited states of the organic fluorophores which are prone to bleaching. Additionally, small spectral shifts were observed more frequently. In our analysis, we only used whole traces. A bleaching event changes the exciton-exciton-interaction, which contaminates the photon statistics simply because the number of physical emitters changes over time. Therefore, it is important that we only sum over traces which arise from the same number of physical emitters that give a constant fluorescence signal over the 10 seconds of recording.

## 2. PL transient of a five-dye sample

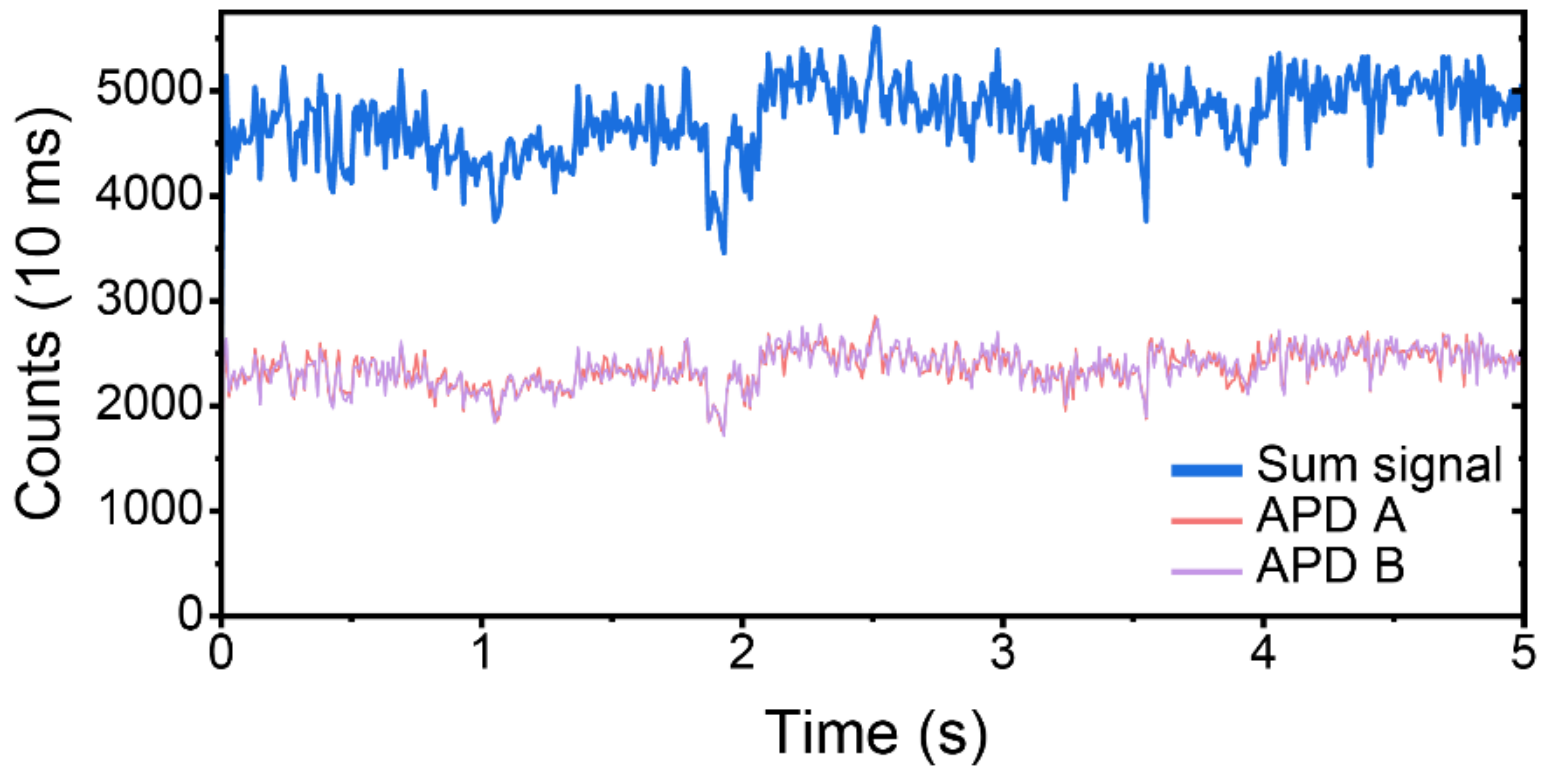

***Figure S3.*** *Representative PL trajectory of a five-dye origami sample with 10 ms binning. The blue trajectory is the sum signal of both detection channels which are shown in red and purple. All five dye trajectories were recorded for five seconds to prevent photodamage and spectral shifts.*

## 3. Instrument response function of the microscope for the DNA origami samples

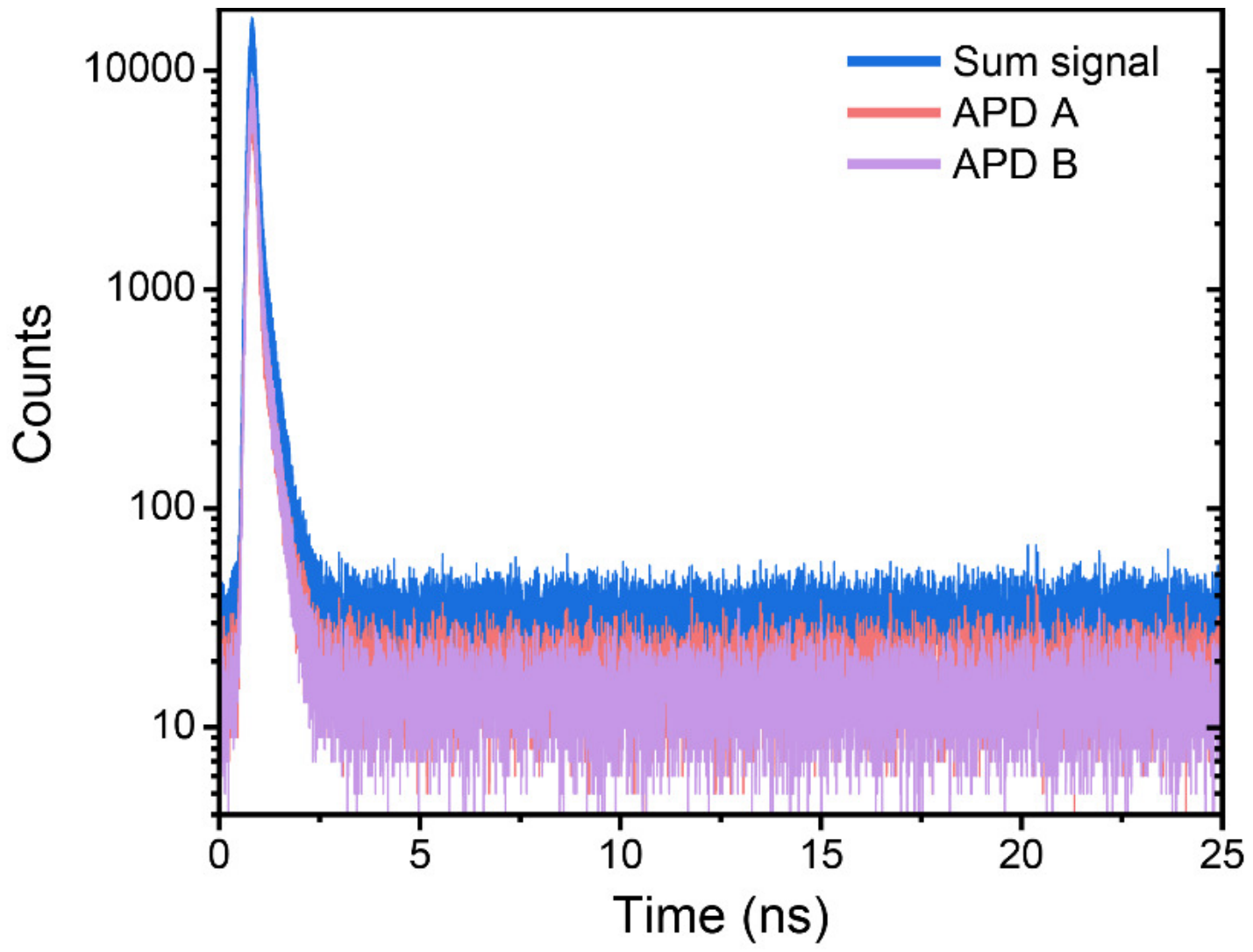

***Figure S4.*** *Instrument response function of the DNA origami setup with 4 ps binning. The microtime = 0 in figure 2b and 3 corresponds to the peak of the sum signal (blue).*

## 4. Number of independent emitters $n$ as a function of SBR

For calculating the expected $n$ for a single dye on the DNA origami setup the signal-to-background ratio (SBR) was calculated from a single-particle trajectory and a background trace from the same sample. Uncorrelated background adds correlation events equally to $N_c$ and $N_l$. This raises the $N_c/N_l$ ratio and therefore a high SBR is desirable. Figure S5a shows the average of the SBR ratios of both APDs. The microtime = 0 corresponds to the peak of the IRF as shown in Figure S4. Due to scattering from the excitation laser pulse the SBR rises after the IRF. After reaching its maximum it decays due to the exponential fluorescence decay. The expected $N_c/N_l$ ratio was calculated according to the adapted relation from Weston *et al.* [5]:

$$\frac{N_c}{N_l} = \frac{1+\frac{S}{B}}{\left(1+\frac{S}{2B}\right)^2} \quad \text{eq. (S1)}$$

This relation gives the expected $N_c/N_l$ ratio for a single independent emitter as a function of signal $S$ and background $B$. For infinite SBR this value becomes zero.

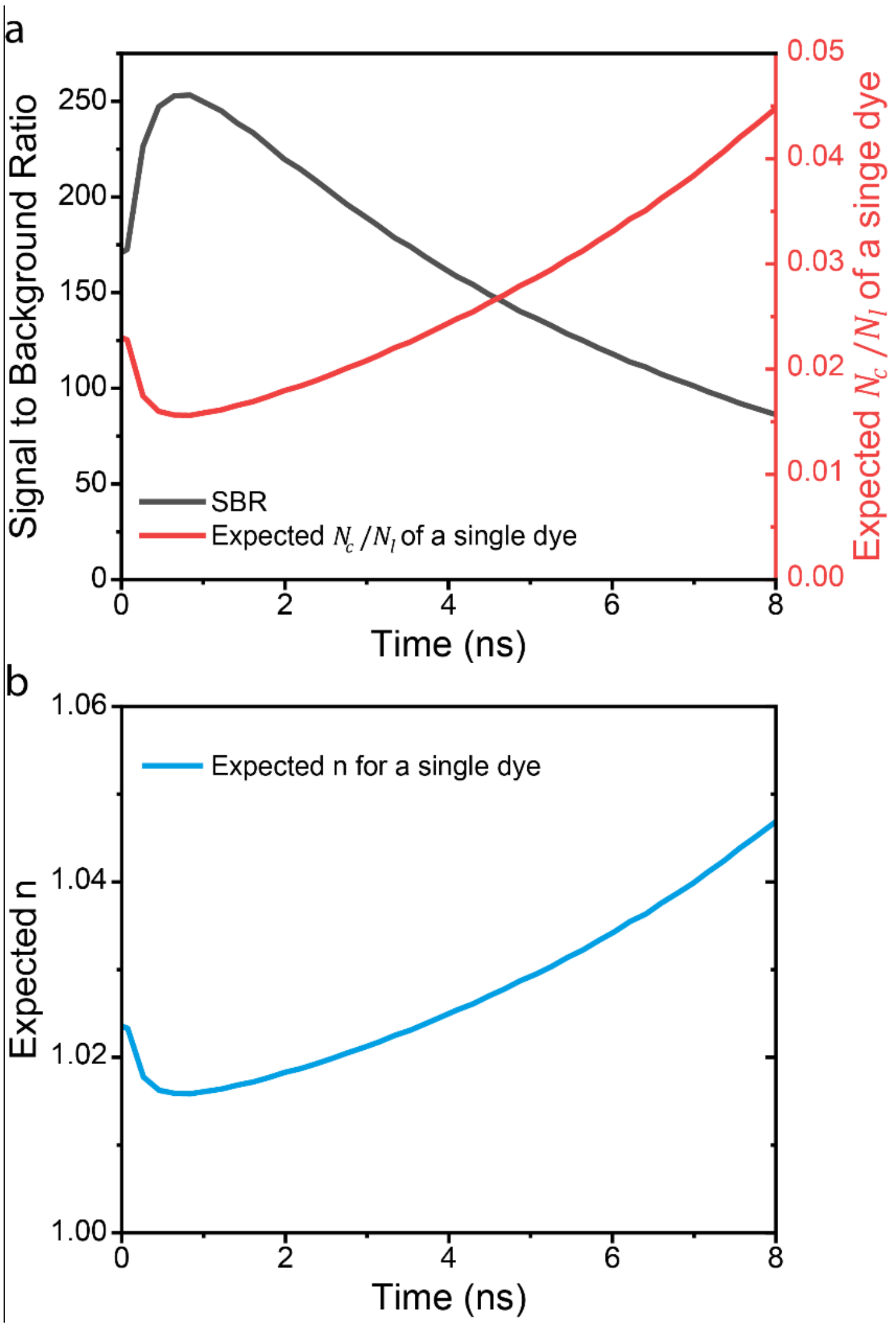

***Figure S5.*** *(a) Signal-to-background ratio of a single fluorescent dye in a DNA origami structure shown in black. Time = 0 corresponds to the peak of the IRF as shown in Figure S4. The corresponding $N_c/N_l$ ratio is shown in red. (b) Expected n for measured SBR in (a) according to eq. (S1). For infinite SBR this value becomes one.*

## 5. Complete fitting results of DNA origami structures

**Table S2:** Complete fitting results with standard error of DNA origami structures according to eq. (3). Values in parentheses were fixed.

| | 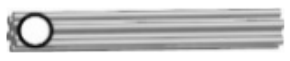 | 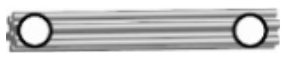 | 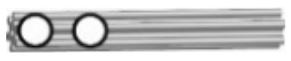 | 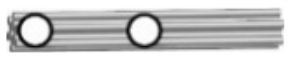 | 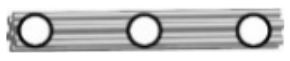 | 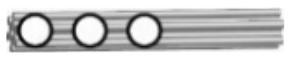 | 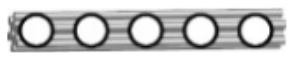 |
|---|---|---|---|---|---|---|---|
| **$y_0$ (1)** | 0.9821 ± 0.0017 | 0.5390 ± 0.0044 | (0.9821) | (0.9821) | (0.9821) | (0.9821) | (0.9821) |
| **$A_1$ (1)** | (0) | 0.0001 ± 0.0043 | 0.4094 ± 0.0057 | 0.414 ± 0.010 | 0.5948 ± 0.0064 | 0.351 ± 0.037 | 0.350 ± 0.037 |
| **$k_1$** ($ns^{-1}$) | (0) | -0.0078 ± 0.0048 | 1.718 ± 0.054 | 0.054 ± 0.0061 | 0.0628 ± 0.0033 | 1.44 ± 0.13 | 1.26 ± 0.12 |
| **$A_2$ (1)** | (0) | (0) | (0) | (0) | (0) | 0.264 ± 0.038 | 0.420 ± 0.038 |
| **$k_2$** ($ns^{-1}$) | (0) | (0) | (0) | (0) | (0) | 0.356 ± 0.038 | 0.272 ± 0.021 |

## 6. Determination of the size of aggregates

Polymer chain numbers are estimated by counting spots in fluorescence images of samples before and after solvent vapour annealing (SVA). The starting concentration allows single polymer chains to be counted, as shown in Figure S6, where we count the spots. This gives us the number of single chains.

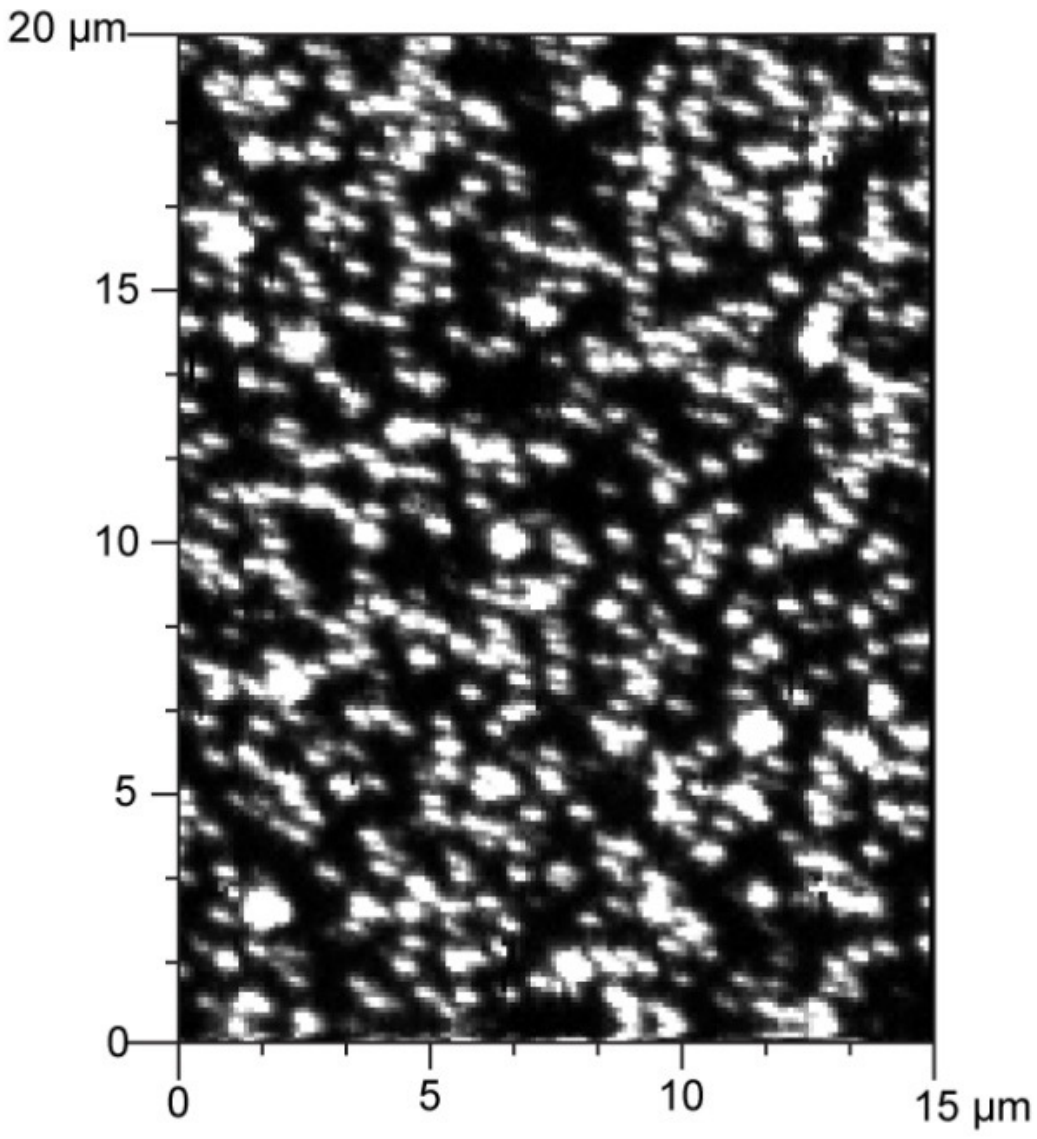

7.

***Figure S6.*** *Fluorescence image of PPEB-2 chains, showing single-chain density.*

We then SVA this film under varying conditions, which swells it, allowing the chains to become mobile and aggregate. We then record a fluorescence image again and count the number of spots, which gives us the number of aggregates as shown in Figure S7.

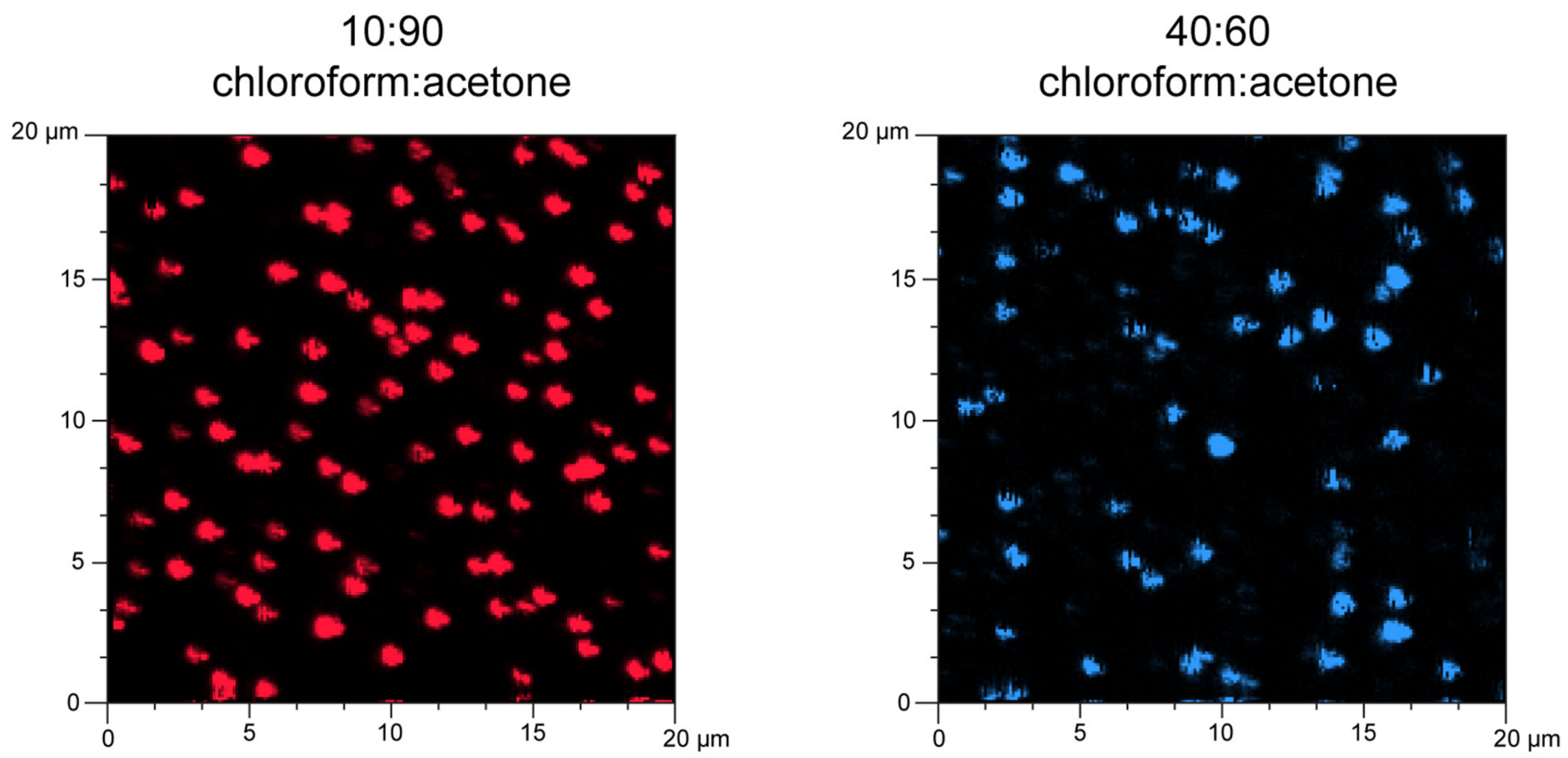

***Figure S7.*** *Fluorescence image of PPEB-2 aggregates after solvent vapour annealing, with chloroform:acetone percentages as denoted. Counting of the number of aggregates across multiple images allows the number of chains per aggregate to be deduced.*

Simple division allows us to ascertain than on average each aggregate thus contains a certain number of chains. We repeat this across several images and under the different SVA conditions to allow us to determine the number of chains per aggregate, with results as shown below in Table S3.

**Table S3:** Average number of PPEB-2 chains per aggregate for two SVA conditions.

| SVA conditions (chloroform:acetone ratio) | Average number of chains per aggregate | Standard deviation |
|---|---|---|
| 10:90 | 8.83 | 0.32 |
| 40:60 | 6.44 | 0.19 |

Knowing the average number of polymer chains contained in an aggregate (54 for PPEB-1[6] and the values as denoted above for PPEB-2), the average molecular weight of a chain (40 kDa with PDI of 1.45 for PPEB-1 and 66 kDa with PDI 1.05 for PPEB-2)[7, 8] and assuming[3] a mass density of 1 g $cm^{-3}$ we can then deduce the average volume of an aggregate. In our case, we get $3.59 \times 10^{-18}$ $cm^3$ for 54 chains of PPEB-1, $9.67 \times 10^{-19}$ $cm^3$ for 9 chains and $7.06 \times 10^{-19}$ $cm^3$ for 6 chains of PPEB-2.

## 7. Quantum-statistical description of psTRAB photon correlations

### Definition of the correlation-ratio observable

For $n_{\mathrm{dyes}}$ chromophores emitting into free space without a cavity, the second-order correlation of emitted photons detected at times $t$ and $t + \Delta t$ can be calculated from the second-order correlation of chromophore deexcitation

$$G^{(2)}(t, t + \Delta t) = \sum_{i,j=0}^{n_{\mathrm{dyes}}} \langle a_i^\dagger(t) a_j^\dagger(t + \Delta t) a_j(t + \Delta t) a_i(t) \rangle, \qquad \text{eq. (S2)}$$

where $a_i(t)$ is the operator on the $2^{n_{\mathrm{dyes}}}$ dimensional Hilbert space that destroys the excitation of chromophore $i$, each chromophore being modeled as a two-level system with only ground and excited states. For the case of excitation by laser impulses at regularly spaced times $t_0$, $t_0 \pm T, t_0 \pm 2T, \ldots$ we define the *central-bin* and the *lateral-bin delay-integrated second-order correlation* at time $t_0 < t < t_0 + T$ as

$$G_{\mathrm{c}}^{(2)}(t) = \int_0^{T-t} G^{(2)}(t, t + \Delta t) d(\Delta t) \quad \text{and} \qquad \text{eq. (S3)}$$

$$G_{\mathrm{l}}^{(2)}(t) = \int_T^{2T-t} G^{(2)}(t, t + \Delta t) d(\Delta t), \qquad \text{eq. (S4)}$$

respectively. Finally, the *central-to-lateral-bin second-order correlation ratio* is defined as $G_{\mathrm{c}}^{(2)}(t)/G_{\mathrm{l}}^{(2)}(t) \equiv N_{\mathrm{c}}(t)/N_{\mathrm{l}}(t)$.

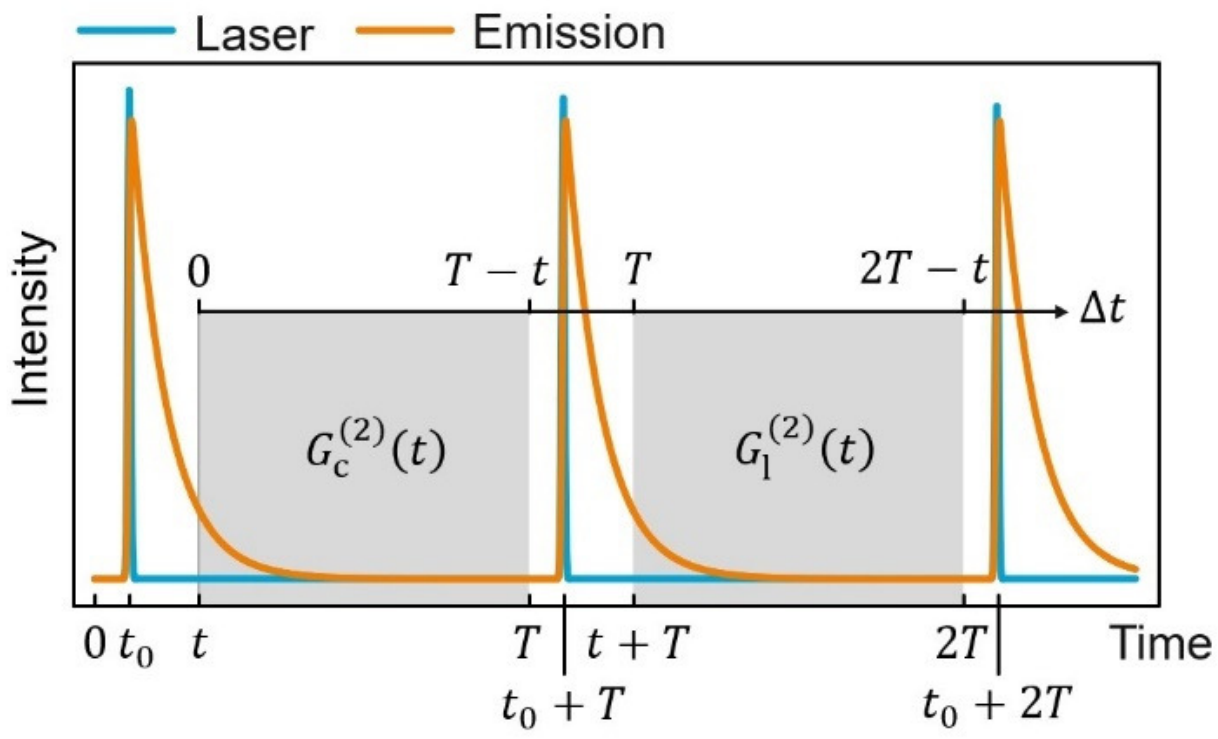

***Figure S8.*** *Visualization of the temporal integration ranges used for $G_{\mathrm{c}}^{(2)}(t)$ and $G_{\mathrm{l}}^{(2)}(t)$. The first photon is detected at time $t > t_0$, and the second photon is detected at time $t + \Delta t$ chosen from either the same central laser repetition period (left gray domain) or the lateral laser repetition period (right gray domain).*

### Quantum-mechanical description of $n_{\mathrm{dyes}}$ coupled chromophores under optical driving

We describe each chromophore by a two-level Hilbert space $\mathcal{H}_2$ with basis states $|0\rangle$ (ground state) and $|1\rangle$ (singlet excited state) and use the usual definitions for the lowering operator $a = |0\rangle\langle 1|$ and the raising operator $a^\dagger = |1\rangle\langle 0|$. The Hamiltonian for a single dye chromophore coupled to a short-impulse laser light field $\mathcal{E}(t) = \mathcal{E}_0 \sin(\omega t) \sum_{j \in \mathbb{Z}} e^{-2\ln(2)\left(\frac{t + j \cdot T}{\Delta t_{\mathrm{fwhm}}}\right)^2}$ is

$$H = \hbar\omega_0\, a^\dagger a + \hbar\Omega(t)\left(a + a^\dagger\right)\sin(\omega t), \qquad \text{eq. (S5)}$$

where $\hbar\omega_0$ is the excited-state energy, $\Omega(t) = \frac{\mu}{\hbar}\mathcal{E}_0 \sum_{j\in\mathbb{Z}} e^{-2\ln(2)\left(\frac{t+jT}{\Delta t_{\text{fwhm}}}\right)^2}$ is the slowly-varying Rabi frequency and $\mu$ is the transition dipole strength. Using the Hermitian operator $A = \omega\, a^\dagger a$ and the corresponding unitary transformation

$$U = e^{iAt} = e^{i\omega t} a^\dagger a + a\, a^\dagger, \qquad \text{eq. (S6)}$$

the Hamiltonian can be transformed as

$$\widetilde{H} = UHU^\dagger + i\hbar \frac{dU}{dt} U^\dagger = UHU^\dagger - \hbar A, \qquad \text{eq. (S7)}$$

resulting in the rotating-frame Hamiltonian

$$\widetilde{H} = \hbar(\omega_0 - \omega)a^\dagger a + \frac{\hbar\Omega(t)}{2i}\left(a\left(1 - e^{-2i\omega t}\right) - a^\dagger(1 - e^{2i\omega t})\right). \qquad \text{eq. (S8)}$$

In the rotating-frame approximation and for resonant excitation $\omega = \omega_0$, high-frequency components at $\pm 2\omega$ are dropped and the Hamiltonian reduces to

$$\widetilde{H} = \frac{\hbar\Omega(t)}{2i}\left(a - a^\dagger\right). \qquad \text{eq. (S9)}$$

Generalizing for the case of $n_{\text{dyes}}$ chromophores, the Hilbert space is constructed as the tensor product $\mathcal{H}_2 \otimes \ldots \otimes \mathcal{H}_2$ and

$$\widetilde{H} = \frac{\hbar\Omega(t)}{2i}\sum_i\left(a_i - a_i^\dagger\right). \qquad \text{eq. (S10)}$$

Since we are not interested in the coherences, calculation of the system dynamics in the rotating frame is sufficient to derive expectation values for the relevant photon correlation. In order to account for spontaneous emission and singlet-singlet annihilation, we describe the system state by a $2^{n_{\text{dyes}}} \times 2^{n_{\text{dyes}}}$ dimensional density matrix $\rho(t)$ and treat its time dynamics on the level of a Lindblad master equation[9]

$$\frac{d}{dt}\rho(t) = \mathcal{L}(t)\rho(t) = -\frac{i}{\hbar}[H(t), \rho(t)] + \sum_{j=1}^{n_{\text{dyes}}} \frac{1}{2}\left(2c_j\rho(t)c_j^\dagger - \rho(t)c_j^\dagger c_j - c_j^\dagger c_j \rho(t)\right) + \sum_{j\neq k}\frac{1}{2}\left(2d_{jk}\rho(t)d_{jk}^\dagger - \rho(t)d_{jk}^\dagger d_{jk} - d_{jk}^\dagger d_{jk}\rho(t)\right). \qquad \text{eq. (S11)}$$

The quantum-jump operators for spontaneous emission are $c_j = \sqrt{k_{\text{r}}}a_j$ and those for singlet-singlet annihilation are $d_{jk} = \sqrt{k_{\text{ET}}}a_j a_k^\dagger a_k$. In the context of our constrained two-level description, the latter operator effectively models energy transfer from chromophore $j$ to chromophore $k$ at a rate $k_{\text{ET}} = \frac{1}{2}k_{\text{SSA}}$, assuming that after excursions to higher excited states (not included in our Hilbert space), the receiving chromophore immediately returns to the first excited state. Here, $k_{\text{SSA}}$ is the total exponential decay rate at which an individual pair of excitations decays by singlet-singlet annihilation to a single excitation.

Two-time correlations for the excitations (i.e. for the emitted photons) can be calculated in the context of the validity of the Lindblad equation by using the quantum-regression theorem[10]

$$\langle a_i^\dagger(t)a_j^\dagger(t+\tau)a_j(t+\tau)a_i(t)\rangle = \text{Tr}\left(a_j^\dagger a_j \Lambda_i(t, t+\tau)\right), \qquad \text{eq. (S12)}$$

where $\Lambda_i(t, t+\tau)$ describes the system that at time $t$ is reduced by one excitation on chromophore $i$ and then follows the same differential equation as the original density matrix.

$$\Lambda_i(t, t) = a_i\rho(t)a_i^\dagger \qquad \text{eq. (S13)}$$

$$\frac{\partial}{\partial \tau}\Lambda_i(t, t+\tau) = \mathcal{L}(t+\tau)\Lambda_i(t, t+\tau). \qquad \text{eq. (S14)}$$

The two-time correlation can thus be calculated by the product of probabilities of removing one excitation from the system at time $t$, and removing the other one at time $t+\tau$ from the system that was disturbed at time $t$ by reducing the number of excitations by one.

As an example, the system dynamics is solved by numerical integration for the case of three chromophores on the *Mathematica* computer algebra platform (Wolfram Research, USA). Parameters where chosen to model chromophores with unity quantum efficiency, a radiative rate of $k_\mathrm{r} = 0.25 \cdot 10^9\ \mathrm{s}^{-1}$, and a singlet-singlet annihilation rate of $k_\mathrm{SSA} = 0.25 \cdot 10^9\ \mathrm{s}^{-1}$. Excitation conditions were $\Delta t_\mathrm{fwhm} = 50$ ps laser impulse width, $T = 50$ ns repetition interval and an instantaneous peak Rabi frequency of $2 \cdot 10^9\ \mathrm{s}^{-1}$. The number of excitations in the coupled set of chromophores can be calculated as $\mathrm{Tr}(\rho(t)\sum_i a_i^\dagger a_i)$.

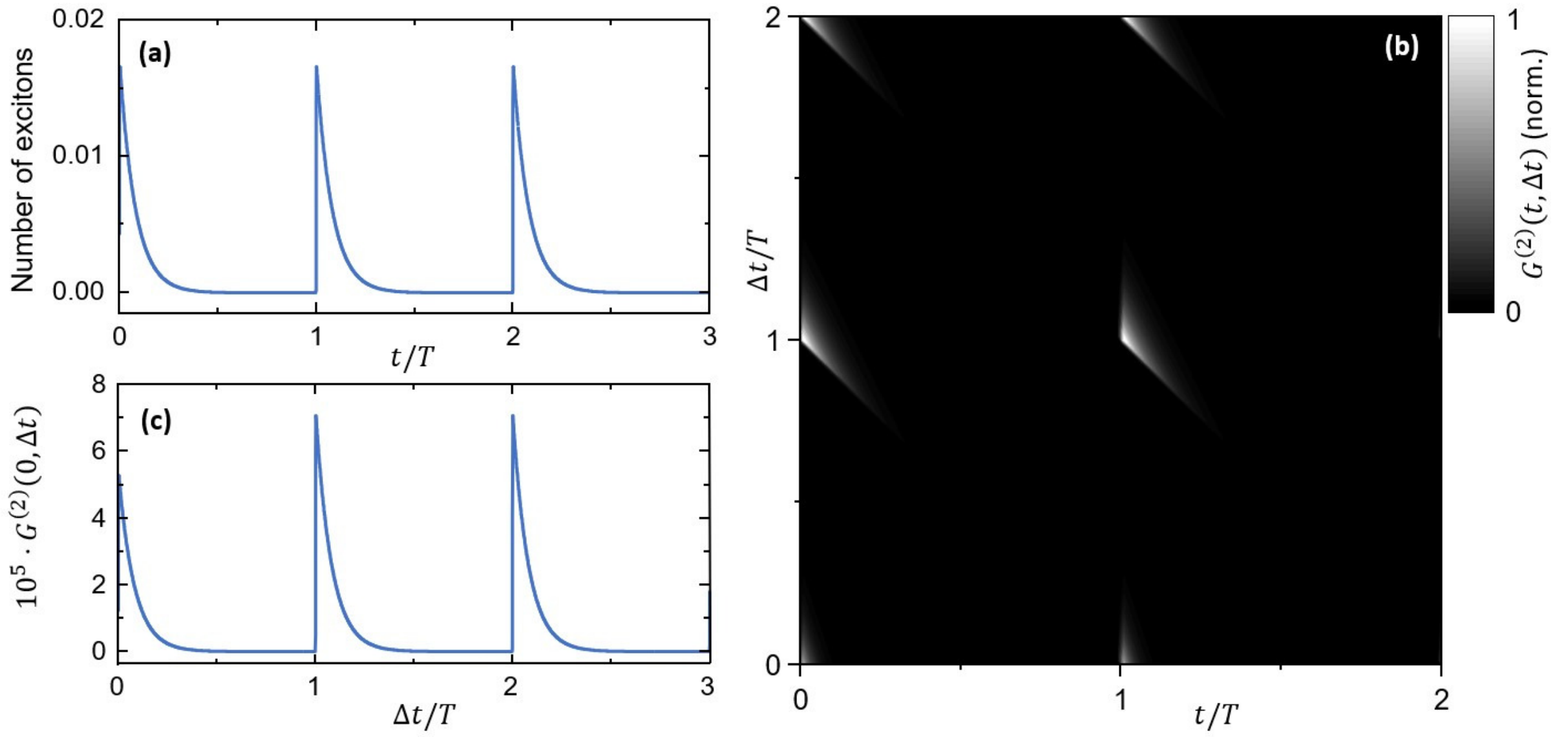

***Figure S9.*** *(a) Number of excitons in a three-chromophore system as a function of time t under pulsed laser excitation. Due to the low excitation strength, singlet-singlet annihilation has no measurable impact on the emission intensity. (b) Normalized two-time correlation $G^{(2)}(t, t+\Delta t)$ as a function of $t$ and $\Delta t$ in units of the pulse repetition period $T$. (c) For delay times $\tau < t - T$, the correlation is reduced due to the loss of one exciton in the system at time $t$. The example shows the case of $t = 0$.*

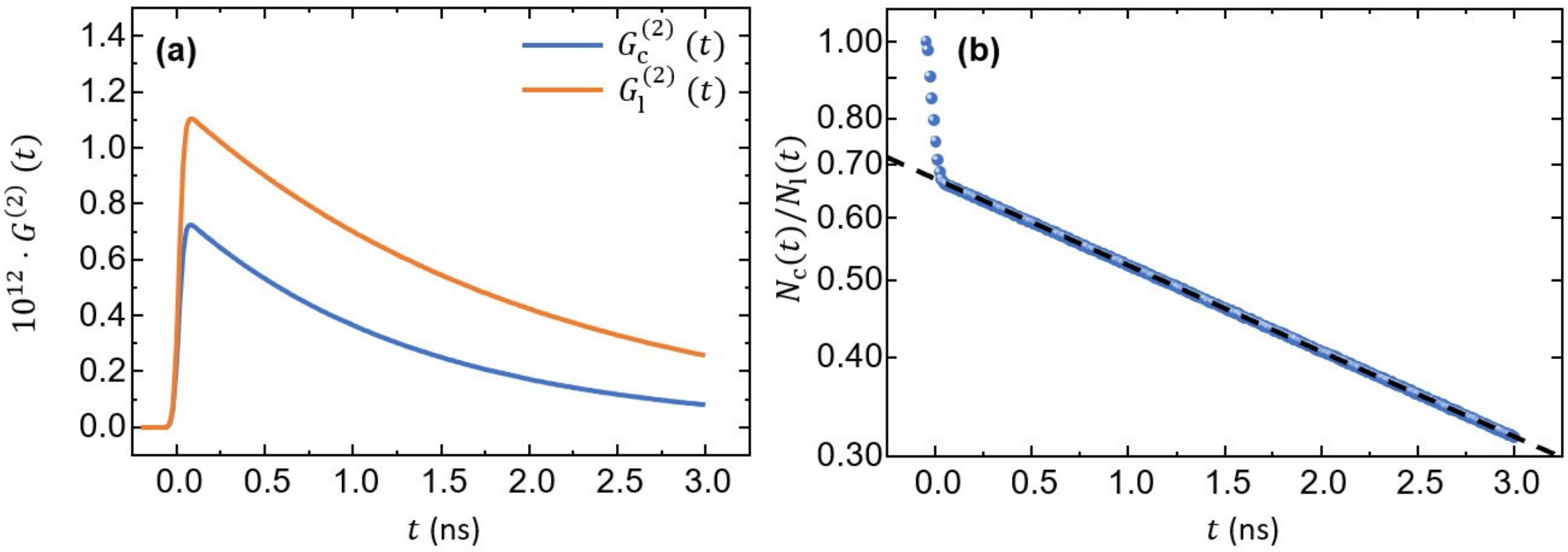

***Figure S10.*** *(a) Central-bin photon correlation* $G_{\mathrm{c}}^{(2)}(t)$ *(blue) and lateral-bin photon correlation* $G_{\mathrm{l}}^{(2)}(t)$ *(orange) as a function of time* t *after laser excitation. (b) After the end of the excitation pulse, the correlation ratio* $N_{\mathrm{c}}(t)/N_{\mathrm{l}}(t) = G_{\mathrm{c}}^{(2)}(t)/G_{\mathrm{l}}^{(2)}(t)$ *(blue) follows the exponential decay* $\frac{n_{\mathrm{dyes}}-1}{n_{\mathrm{dyes}}} e^{-k_{\mathrm{SSA}}t}$ *that directly reflects the annihilation dynamics of exciton pairs (dashed line). The initial decay of the correlation ratio from 1 to* $(n_{dyes} - 1)/n_{\mathrm{dyes}}$ *is due to the high probability of re-excitation if the first photon is detected early during the excitation pulse.*

## Analytical treatment

The system can be treated analytically for the case of negligibly short excitation impulses, assuming that the dynamics is treated in the limit of vanishing Rabi frequency $\Omega \rightarrow 0$. We first discuss the the $n_{\mathrm{dyes}} = 2$ system analytically, assuming that at $t < 0$ the system reaches equilibrium due to optical pumping. In this case, the total number of excitations in the system is $p_0 n_{\mathrm{dyes}}$, where the probability $p_0$ for a single dye chromophore to be excited depends on the excitation intensity. From the probability of detecting a photon we find the emission intensity $I(t) = k_{\mathrm{r}}\mathrm{Tr}\left(\left(a_1^\dagger a_1 + a_2^\dagger a_2\right)\rho(t)\right)$ to be

$$I(t) = \frac{k_{\mathrm{r}}}{2} e^{-(2k_{\mathrm{r}}+k_{\mathrm{SSA}})t} p_0 n_{\mathrm{dyes}} \left(\frac{k_{\mathrm{SSA}} p_0 n_{\mathrm{dyes}}}{4(k_{\mathrm{r}}+k_{\mathrm{SSA}})} + \left(1 - \frac{k_{\mathrm{SSA}} p_0 n_{\mathrm{dyes}}}{4(k_{\mathrm{r}}+k_{\mathrm{SSA}})}\right) e^{(k_{\mathrm{r}}+k_{\mathrm{SSA}})t}\right) \quad \text{eq. (S15)}$$

Experimentally, one works in the low-excitation regime $p_0 n_{\mathrm{dyes}} \ll 1$, and more specifically $k_{\mathrm{SSA}} p_0 n_{\mathrm{dyes}} \ll 4(k_{\mathrm{r}} + k_{\mathrm{SSA}})$. In this limit, the intensity is $I(t) \approx k_{\mathrm{r}} p_0 e^{-k_{\mathrm{r}}t}$. Since in this case the system is only ever singly excited, the photon correlations are calculated as

$$G_{\mathrm{c}}^{(2)}(t, t+\tau) = k_{\mathrm{r}}^2 n_{\mathrm{dyes}}(n_{\mathrm{dyes}} - 1)\mathrm{Tr}\left(a_2^\dagger a_2\, \Lambda(t, t+\tau)\right) \cdot \mathrm{Tr}(a_1^\dagger a_1 \rho(t)) \quad \text{eq. (S16)}$$

$$G_{\mathrm{l}}^{(2)}(t, t+\tau) = k_{\mathrm{r}}^2 n_{\mathrm{dyes}}^2 \mathrm{Tr}\left(a_2^\dagger a_2\, \rho(t+\tau)\right) \cdot \mathrm{Tr}\left(a_1^\dagger a_1 \rho(t)\right). \quad \text{eq. (S17)}$$

Since the analytical description lacks the action of the pulsed excitation, the central-bin and lateral-bin correlations must be calculated differently than in the experimental case. For the central-bin correlation, annihilation of one excitation at time $t$ on chromophore 1 defines the new initial state $\Lambda(t, t) = a_1 \rho(t) a_1^\dagger$. There are $n_{\mathrm{dyes}} = 2$ choices for the first emitting chromophore and only $(n_{\mathrm{dyes}} - 1) = 1$ choice for the second, since the first chromophore cannot be reexcited. For the lateral-bin correlation, annihilation of the first exciton on one of the $n$ chromophores does not change the density matrix, since the system fully recovers from one to the next excitation pulse. The number of available chromophores to emit the second photon is thus also $n_{\mathrm{dyes}} = 2$. In the limit of $p_0 n_{\mathrm{dyes}} \ll 1$, one finds

$$G_{\mathrm{c}}^{(2)}(t, t+\tau) = \frac{k_{\mathrm{r}}^2}{4} n_{\mathrm{dyes}}(n_{\mathrm{dyes}} - 1)\, p_0^2 e^{-(k_{\mathrm{r}}+k_{\mathrm{ET}})t} e^{-(k_{\mathrm{r}}+k_{\mathrm{ET}})t} e^{-k_{\mathrm{r}}\tau} \quad \text{eq. (S18)}$$

$$G_{\mathrm{l}}^{(2)}(t, t+\tau) = \frac{k_{\mathrm{r}}^2}{4} n_{\mathrm{dyes}}^2 p_0^2 e^{-k_{\mathrm{r}}t} e^{-k_{\mathrm{r}}t} e^{-k_{\mathrm{r}}\tau} \quad \text{eq. (S19)}$$

Integration over the delay time $\tau$ then yields

$$G_{\mathrm{c}}^{(2)}(t) = \int_0^\infty G_{\mathrm{c}}^{(2)}(t, t+\tau) d\tau = \frac{k_{\mathrm{r}}}{4} n_{\mathrm{dyes}}(n_{\mathrm{dyes}} - 1)\, p_0^2 e^{-(2k_{\mathrm{r}}+k_{\mathrm{SSA}})t} \quad \text{eq. (S20)}$$

$$G_{\mathrm{l}}^{(2)}(t) = \int_0^\infty G_{\mathrm{l}}^{(2)}(t, t+\tau) d\tau = \frac{k_{\mathrm{r}}}{4} n_{\mathrm{dyes}}^2 p_0^2 e^{-2k_{\mathrm{r}}t} \quad \text{eq. (S21)}$$

and a correlation ratio of

$$\frac{N_\mathrm{c}(t)}{N_\mathrm{l}(t)} = \frac{G_\mathrm{c}^{(2)}(t)}{G_\mathrm{l}^{(2)}(t)} = \frac{n_\mathrm{dyes}-1}{n_\mathrm{dyes}} e^{-k_\mathrm{SSA}t}. \quad \text{eq. (S22)}$$

These results for $n_\mathrm{dyes} = 2$ can be directly rationalized for higher numbers of chromophores and the case of non-radiative decay at rate $k_\mathrm{nr}$. Under conditions of $p_0 n_\mathrm{dyes} \ll 1$, the probability to find two excited chromophores after each of two adjacent excitation pulses is negligible, even for systems with $n_\mathrm{dyes} > 2$ chromophores. Thus, lateral-bin correlations derive from chromophore ensembles with only *one* excitation present and are devoid of the effect of singlet-singlet annihilation. On the other hand, central-bin correlations can only exist if indeed two or more excitations are present after the laser excitation pulse. Since $p_0 n_\mathrm{dyes}$ is small, cases with more than two excitations need not be considered. Obviously, the probability to find two excitations decays proportionally to $e^{-k_\mathrm{SSA}t}$, where $k_\mathrm{SSA}$ is the *exponential decay rate for pairs of singlet excitons*, such that in a macroscopic ensemble with a high number of excitons $n_\mathrm{exc}$, one would expect a non-exponential decay according to $\frac{d}{dt} n_\mathrm{exc} = -n_\mathrm{exc}(n_\mathrm{exc}-1)k_\mathrm{SSA}$. Microscopically however, each excited chromophore is exponentially quenched by each other excited chromophore at the *energy transfer rate* $k_\mathrm{ET} = k_\mathrm{SSA}/2$. Applying the quantum-regression theorem, the probability of finding the second photon emission at time $t+\tau$ for the central-bin second-order correlation is calculated by assuming that the number of excitons is *reduced to one* at time $t$. There is a total number of $n$ chromophores to choose from as source of the first detected photon, and neglecting the possibility of double-excitations of individual chromophores, a total of $(n_\mathrm{dyes} - 1)$ chromophores exist that can potentially carry the second exciton. The central-bin second-order correlation is thus calculated as

$$G_\mathrm{c}^{(2)}(t, t+\tau) = n_\mathrm{dyes} k_\mathrm{r}\, p_0\, e^{-(k_\mathrm{r}+k_\mathrm{nr}+k_\mathrm{ET})t} \cdot k_\mathrm{r}(n_\mathrm{dyes}-1) p_0 e^{-(k_\mathrm{r}+k_\mathrm{nr}+k_\mathrm{ET})t} e^{-(k_\mathrm{r}+k_\mathrm{nr})\tau}$$

eq. (S23)

$$G_\mathrm{c}^{(2)}(t) = n_\mathrm{dyes}(n_\mathrm{dyes}-1) p_0^2 k_\mathrm{r}^2 \tau_\mathrm{PL} e^{-2(k_\mathrm{r}+k_\mathrm{nr})t} e^{-k_\mathrm{SSA}t}, \quad \text{eq. (S24)}$$

where $\tau_\mathrm{PL} = 1/(k_\mathrm{r} + k_\mathrm{nr})$ is the single-chromophore photoluminescence lifetime. For the lateral-bin correlation, the only relevant contribution comes from the configuration where there is exactly one exciton in the chromophore system when the first photon is emitted and exactly one photon in the chromophore system when the second photon is emitted after the next (lateral) excitation pulse. Thus, singlet-singlet quenching does not contribute to the lateral-bin second-order correlation. Again, $n_\mathrm{dyes}$ chromophores can potentially emit the first photon, but in contrast to the calculation of the central-bin correlation, for the lateral-bin correlation the system is re-excited in-between the two photon detection events, and thus the number of chromophores that can emit the second photon is also $n_\mathrm{dyes}$:

$$G_\mathrm{l}^{(2)}(t, t+\tau) = n_\mathrm{dyes} k_\mathrm{r}\, p_0\, e^{-(k_\mathrm{r}+k_\mathrm{nr})t} \cdot k_\mathrm{r} n_\mathrm{dyes}\, p_0 e^{-(k_\mathrm{r}+k_\mathrm{nr})t} e^{-(k_\mathrm{r}+k_\mathrm{nr})\tau} \quad \text{eq. (S25)}$$

$$G_\mathrm{l}^{(2)}(t) = n_\mathrm{dyes}^2 p_0^2 k_\mathrm{r}^2 \tau_\mathrm{PL} e^{-2(k_\mathrm{r}+k_\mathrm{nr})t}. \quad \text{eq. (S26)}$$

The resulting central-to-lateral-bin second-order correlation ratio for the general case of $n_\mathrm{dyes}$ chromophores is thus again simply

$$\frac{N_\mathrm{c}(t)}{N_\mathrm{l}(t)} = \frac{G_\mathrm{c}^{(2)}(t)}{G_\mathrm{l}^{(2)}(t)} = \frac{n_\mathrm{dyes}-1}{n_\mathrm{dyes}} e^{-k_\mathrm{SSA}t}. \quad \text{eq. (S27)}$$

An important observation in this treatment is that the choice of the singlet-singlet annihilation mechanism and the associated decay law $e^{-k_\mathrm{SSA}t}$ is entirely arbitrary. The immediate conclusion is that the correlation ratio can be used to directly determine the actual decay function for the doubly-excited system.

**Table S4:** Sequences for the modified DNA origami structure.

| 5‘ position | Sequence | Comment |
|---|---|---|
| 0[286] | AAAACGAAAGAGGCTCATTATAC | |
| 11[105] | ACACAACATACGAGGGATGTGGCTATTAATCGGCC | |
| 9[567] | ATCATTTACATAAAAGTATCAAAATTATAAGAAACTTCAATA | |
| 7[567] | CAGCTTTGAATACCAAGTTACAA | |
| 5[455] | CATGCCAGTGAGCGCTAATATCCAATAATAAGAGC | |
| 2[223] | CCGAACTTTAATAAAAGCAAAGCGGATT | |
| 5[497] | TTGAGAATATCTTTCCTTATCACTCATCGAGAACA | |
| 9[315] | CAGATATAGGCTTGAACAGACGTTAGTAAAGCCCAAAAATTT | |
| 5[287] | GCGCAGCGACCAGCGATTATATATCATCGCCTGAT | |
| 8[69] | TCGGTCATACCGGGGGTTTCTGC | |
| 11[219] | GTGCCTGCTTTAAACAGGGAGAGAGTTTCAAAGCGAACCA | |
| 10[457] | AAAAGATAGGGTTGAGTGT | |
| 2[643] | GATAGTGCAACATGATATTTTTGAATGG | |
| 0[347] | AGCGTATCATTCCACAGACCCGCCACAGTTGCAGCAAGCG | |
| 9[483] | ATAATGAATCCTGAGATTACGAGCATGTGACAAAAACTTATT | |
| 8[573] | AAATGCGTTATACAAATTCTTAC | |
| 2[433] | AGGGACAAAATCTTCCAGCGCCAAAGAC | |
| 7[63] | GCCCGCACAGGCGGCCTTTAGTG | |
| 4[377] | CTATTTCGGAACGAGTGAGAATA | |
| 0[698] | TTTTTCGGGAGCTAAACAGGTTGTTAGAATCAGAGTTTTT | |
| 4[587] | CATCGGGAGAAATTCAAATATAT | |
| 7[506] | AAATCAGCCAGTAATAACACTATTTTTGAAGCCTTAAATC | |
| 7[170] | TTTTTATCCAATAAATCTCTACCCCGGTAAAACTAGCATG | |
| 5[161] | GTATACAGGTAATGTGTAGGTAGTCAAATCACCAT | |
| 4[396] | AACAGAGTGCCTGGGGTTTTGCTCACAGAAGGATTAGGAT | |
| 3[350] | GTCACCAGTACAAGGTTGAGGCA | |
| 5[581] | ACATCATTTAAATTGCGTAGAAACAGTACCTTTTA | |
| 5[623] | ATACCCTTCGTGCCACGCTGAACCTTGCTGAACCT | |
| 8[130] | GGGCGTGAAATATTAGCGCCATTCGC | |
| 9[357] | TCTTATACTCAGAAAGGCTTTTGATGATATTGACACGCTATT | |
| 11[345] | GAGAGCCTCAGAACCGCATTTTCTGTAACGATCTAAAGTT | |
| 5[329] | TTCATTTTCTGCTAAACAACTGAACAACTAAAGGA | |
| 8[489] | AAAACGGAATACCCAAAAGAACT | |
| 0[202] | GACCGGAAGCAATTGCGGGAGAA | |
| 3[182] | GCTAAATCGGTTTGACTATTATA | |
| 3[392] | ATATTCACAAACAAATTCATATG | |
| 6[69] | AAAAGTGTCAGCAACAATTGCAGGCGCT | |
| 11[567] | ACCATCACCCAAATAAACAGTTCATTTGATTCGCC | |
| 7[590] | AATCGTTGAGTAACATTGGAATTACCTAATTACATTTAAC | |
| 11[93] | GCTCAAGTTGGGTAACGGGCGGAAAAATTTGTGAGAGATA | |
| 0[305] | ACTACCTTTAAACGGGTAACAGGGAGACGGGCA | |
| 4[270] | TCAACATCAGTTAAATAGCGAGAGTGAGACGACGATAAAA | |
| 6[153] | TAAATCGGTTGGTGCACATCAAAAATAA | |
| 10[163] | TCAGCTAACTCACATTAAT | |
| 7[231] | TGCAACACTATCATAACCCTCGT | |
| 4[438] | ACCAAATTACCAGGTCATAGCCCCGAGTTTTCATCGGCAT | |
| 8[195] | TTAACAAGAGAATCGATGAACGG | |
| 3[625] | AGACAACCTGAACAGTATTCGAC | |
| 8[363] | TGAACAGCTTGATACCGATAGTT | |
| 0[412] | TCACCGTCACCGGCGCAGTCTCT | |
| 4[706] | TTTTTGTCCATCACGCAAATTCCGAGTAAAAGAGTCTTTTTT | |
| 11[315] | ACAGCTGATTGCCCGTCGCTGCGCCCACACGTTGA | |
| 8[424] | CGGAAGCACGCAAACTTATTAGCGTT | |
| 0[431] | ATTCAAGGGGAAGGTAAATGTGGCAAATAAATC | |
| 3[602] | TGATTATCAGATATACGTGGCAC | |
| 4[545] | TGACCTAAATTTTTAAACCAAGT | |
| 3[679] | GGTTGCTTTGACGAGCACGTTTTT | |
| 6[573] | TGATTTAGAAAACTCAAGAGTCAATAGT | |
| 11[441] | AAAAGAATAGCCCGATACATACGCAGTAAGCTATC | |
| 8[634] | TACATAAATTCTGGGCACTAACAACT | |
| 3[541] | CATAGTTAATTTGTAAATGTCGC | |
| 11[147] | TGCCTAATGAGTGAGAAAAGCTCATATGTAGCTGA | |
| 9[651] | AATAGCTGTCACACGCAACGGTACGCCAGCGCTTAATGTAGTA | |
| 0[557] | TACCTAATATCAAAATCATTCAATATTACGTGA | |
| 4[60] | TCAGAGGTGTGTCGGCCAGAATGAGTGCACTCTGTGGT | |
| 3[476] | TTTTTTGTTTAATAAAGTAATTC | |
| 8[382] | AAGTAAGAGCCGCCAGTACCAGGCGG | |
| 3[79] | GTGGAACGACGGGCTCTCAACTT | |
| 4[102] | CCAGCCAAACTTCTGATTGCCGTTTTGGGTAAAGTTAAAC | |
| 3[121] | AATCAGTTAAAACGTGGGAGAAA | |
| 3[224] | GCATCAAAAAGAAGTAAATTGGG | |
| 7[212] | TTTCACGAGAATGACCATTTTCATTTGGTCAATAACCTGT | |
| 8[678] | CCTACATACGTAGCGGCCAGCCATTGCAACAGGTTTTT | |
| 5[539] | TTCGCTATTCGCAAGACAAAGTTAATTTCATCTTC | |
| 7[17] | TTTTTATCCAGCGCAGTGTCACTGC | |
| 8[298] | CATAGAATTTGCGGTTTGAAAGAGGA | |
| 10[79] | GTATGTGAAATTGTTATCC | |
| 7[273] | ACTACTTAGCCGGAACGAGGCGC | |
| 11[387] | GGCGACACCACCCTCAGGTTGTACTGTACCGTTCCAGTAA | |
| 6[447] | TTACCTCTTAGCAAATTTCAACCGATTG | |
| 8[508] | GGTTTGCGCATTTTAACGCGAGGCGT | |
| 10[415] | CCTCCGAAATCGGCAAAAT | |
| 4[480] | TAAGCCAGAGAGCCAGAAGGAAACTCGATAGCCGAACAAA | |
| 0[179] | GCCTTATACCCTGTAATACCAATTCTTGCGCTC | |
| 9[147] | CATTCAACCCAAAATGTAGAACCCTCATGAATTAGTACAACC | |
| 7[525] | TATGTGATAAATAAGGCGTTAAA | |
| 0[454] | AGACGGGAGAATTGACGGAAATT | |
| 11[681] | AAAGGGCGCTGGCAAGTATTGGC | |
| 4[228] | GAGCTTAAGAGGTCCCAATTCTGCAATTCCATATAACAGT | |
| 3[331] | TACCGGGATAGCAATGAATATAT | |
| 4[335] | ATTGCGAATAATGTACAACGGAG | |
| 2[265] | TATGCATTACAGAGGATGGTTTAATTTC | |
| 4[564] | TTTAGAACGCGAATTACTAGAAAACTATAAACACCGGAAT | |
| 11[597] | GAGGTAACGTTATTAATTTTAAAACAAATAATGGAAGGGT | |
| 5[25] | TTTTTCCGGTGCAGCACCGATCCCTTACACTTGCC | |
| 1[17] | TTTTTTGGTAATGGGTAACCATCCCACTTTTT | |
| 8[531] | AACGAACCTCCCGACTTGCGGGA | |
| 0[515] | CTGAAAACCTGTTTATCAAACATGTAACGTCAA | |
| 8[592] | AAAATTTTTTAAAATGAGCAAAAGAA | |
| 7[609] | ATTTGGCAAATCAACAGTTGAAA | |
| 11[639] | CCGATAATAAAAGGGACTTAACACCGCGAACCACCAGCAG | |
| 3[583] | GGAATCGGAACATTGCACGTTAA | |
| 2[349] | TGTAGGGGATTTAGTAACACTGAGTTTC | |
| 3[434] | AAAAGGGCGACAATTATTTATCC | |
| 5[371] | ATCAGAGCCTTTAACGGGGTCTTAATGCCCCCTGC | |
| 7[338] | GGAGCAGCCACCACCCTTCGCATAACGACAATGACAACAA | |
| 3[56] | ATCAGCGGGGTCAGCTTTCAGAG | |
| 0[473] | AAAAAAGGCAGCCTTTACAATCTTACCAGTTTG | |
| 6[698] | TTTTTAACAATATTACCGTCGCTGGTAATATCCAGTTTTT | |
| 8[88] | AGCCTCCCCAGGGTCCGGCAAACGCG | |
| 6[405] | CAAGTGCTGAGTAAGAAAATAAATCCTC | |
| 7[632] | GGAATAACAGAGATAGACATACAAACTTGAGGATTTAGAA | |
| 0[76] | GACTTTCTCCGTGGCGCGGTTG | |
| 2[97] | GCGAAAGACGCAAAGCCGCCACGGGAAC | |
| 4[648] | GCATCGAGCCAGATATCTTTAGGACCTGAGGAAGGTTATC | |
| 4[606] | ACAGTTTTTCAGATTTCAATTACCGTCGCAGAGGCGAATT | |
| 7[548] | TAAGATCTGTAAATCGTTGTTAATTGTAAAGCCAACGCTC | |
| 11[555] | CCCACATGTGAGTGAATAACTGATGCTTTTAACCTCCGGC | |
| 9[399] | ATAAGAAGCCACCCAAACTTGAGCCATTATCAATACATCAGT | |
| 11[189] | ACTGCCCGCTTTCCTGAAAAGCTATATTTTAAATA | |
| 3[499] | TGTCCAAGTACCAGAAACCCCAG | |
| 4[209] | AATGCTGTAGCTGAGAAAGGCCG | |
| 7[357] | GTGTATTAAGAGGCTGAGACTCC | |
| 8[237] | GCTTGACCATTAGATACATTTCG | |
| 9[609] | GATGAATAAATCCTGTAGGTGAGGCGGTAGCGTAAGTCCTCA | |
| 0[328] | TTGTCGTCTTTCTACGTAATGCC | |
| 11[513] | CTCCAATTTAGGCAGAGACAATCAATCAAGAAAAATAATA | |
| 3[560] | AAGACGCTGAGACCAGAAGGAGC | |
| 7[42] | GCGCCTGAATGCCAACGGCCCAGCCTCCCGCGTGCCTGTTCTTCTTTTT | |
| 0[370] | GCGTCATACATGCCCTCATAGTT | |
| 11[303] | GTGAGTTAAAGGCCGCTGACACTCATGAAGGCACCAACCT | |
| 3[373] | GGTCACGCCAGCACAGGAGTTAG | |
| 4[51] | GGGTTACCTGCAGCCAGCGGTGTTTTT | |
| 7[254] | TTACCAATAAGGCTTGCAGTGCGGAAGTTTAGACTGGATA | |
| 8[466] | GGCATAAGCGTCTTCGAGGAAACGCA | |
| 3[247] | CTTGAAAACACCCTAACGGCATA | |
| 8[405] | GGTGCCGTCGAGAGGGTTGATAT | |
| 10[331] | TCGTTCACCGCCTGGCCCT | |
| 8[615] | GTTGAAACAAACATCAAGAAAAC | |
| 6[531] | GACCGTCGAACGGGGAAGCTAATGCAGA | |
| 6[363] | TGAAATTGTTTCAGGGAACTACAACGCC | |
| 10[625] | AACACCCTAAAGGGAGCCC | |
| 6[279] | CATGTCAGAGATTTGATGTGAATTACCT | |
| 11[429] | CCCTTCATATAAAAGAACGTAGAGCCTTAAAGGTGAATTA | |
| 11[651] | TTGACGGGGAAAGCTTCACCAGAAATGGCATCACT | |
| 6[615] | GTCAGTCGTTTAACGAGATGGCAATTCA | |
| 7[422] | AGCGCCACCACGGAATACGCCTCAGACCAGAGCCACCACC | |
| 4[312] | ATTTGCCAAGCGGAACTGACCAACGAGTCAATCATAAGGG | |
| 8[550] | CAGTAAGAACCTTGAGCCTGTTTAGT | |
| 4[503] | AGCAAGCCGTTTAAGAATTGAGT | |
| 2[601] | TCAATAATAAAGTGTATCATCATATTCC | |
| 9[21] | TTTTTGCGTCCGTGCCTGCATCAGACGTTTTT | |
| 11[483] | GAACAAGAGTCCACCAATTTTTTAGTTGTCGTAGG | |
| 10[499] | CTATATTAAAGAACGTGGA | |
| 4[186] | GAGACAAAGATTATCAGGTCATTGACGAGAGATCTACAAA | |
| 9[63] | TTCACCTAGCGTGGCGGGTGAAGGGATACCAGTGCATAAAAA | |
| 11[609] | AGCACTAAATCGGATCGTATTTAGACTTATATCTG | |
| 4[293] | AAATTGTGTCGAGAATACCACAT | |
| 3[667] | GGCGCCCCGCCGAATCCTGAGAAGTGAGGCCGATTAAAGG | |
| 3[205] | GTCAGAATCAGGCAGGATTCGCG | |
| 0[622] | AAGATAAAACAGTTGGATTATAC | |
| 6[111] | TCAGGTGAAATTTCTACGGAAACAATCG | |
| 10[205] | AGCAGTCGGGAAACCTGTC | |
| 6[489] | AATCATAATAACCCGGCGTCAAAAATGA | |
| 0[496] | TCCCATCCTAATGAGAATAACAT | |
| 0[221] | CGAGCACAGACTTCAAATACCTCAAAAGCTGCA | |
| 9[231] | TTAGTGTGAATCCCTCTAATAAAACGAAAGAACGATGAATTA | |
| 4[629] | CAAATATCAAACCAGATGAATAT | |
| 0[664] | GATTTTAGACAGGCATTAAAAATA | |
| 10[667] | AGACGGCGAACGTGGCGAG | |
| 0[599] | TTCTGGAATAATCCTGATTTTGCCCGGCCGTAA | |
| 3[23] | TTTGCAACCAGCTTACGGCGGTGGTGAGGTTTCAGTTGAGGATCCTTTTT | |
| 8[340] | GCGCCCGCACCCTCTCGAGGTGAATT | |
| 7[674] | GCCTTACGCTGCGCGTAAAATTATTTTTGACGCTCAATC | |
| 7[86] | ATGAATCCCAGTCACGATCGAACGTGCCGGCCAGAGCACA | |
| 5[245] | CGCCTGACGGTAGAAAGATTCTAATGCAGATACAT | |
| 8[657] | GTATTAGAGCCGTCAATAGATAA | |
| 3[308] | CTAAAGACTTTTAGGAACCCATG | |
| 2[702] | TTTTTTATAACGTGCTTTCCTCTTTATAACAGTACTAT | |
| 4[671] | TACTTCTTTGATAAAAATCTAAA | |
| 2[391] | ATTAAAATAAGTGCGACGATTGGCCTTG | |
| 9[189] | GAGCAAGGTGGCATTTACTCCAACAGGTTCTTTACGTCAACA | |
| 4[167] | CAATATGATATTGATGGGCGCAT | |
| 7[147] | GCTAATGCCGGAGAGGGTAGCTA | |
| 7[464] | AAGCACAGAGCCTAATTATTGTTAGCGATTAAGACTCCTT | |
| 8[172] | TAATCGTAGCATTACCTGAGAGTCTG | |
| 0[580] | TAGAACCTACCAGTCTGAGAGAC | |
| 4[354] | GAAAGTTCAACAATCAGCTTGCTTAGCTTTAATTGTATCG | |
| 8[46] | CAGCATCAACCGCACGGCGGGCCGTT | |
| 2[181] | TTATGGCCTGAGCACCTCAGAGCATAAA | |
| 3[644] | CTATTAGTCTTTCGCCGCTACAG | |
| 8[111] | CTTTTTTTCGTCTCGTCGCTGGC | |
| 11[231] | TTAATGAATCGGCCATTCATTCCAATACGCATAGT | |
| 3[518] | AACAACATGTTCATCCTTGAAAA | |
| 5[77] | AACGTTGTAGAAACAGCGGATAGTTGGGCGGTTGT | |
| 10[706] | TTTTTAGGAGCGGGCGCTAGGAAGGGAAGAAAGCGAATTTTT | |
| 9[441] | TGCCATACATAAAGATTAACTGAACACCAACAGCCGGAATAG | |
| 7[189] | GGCTAAAGTACGGTGTCTGGAAG | |
| 6[237] | AAGAGATTCATTTTGTTTAAGAGGAAGC | |
| 5[203] | TGTAAATCATGCTCCTTTTGATAATTGCTGAATAT | |
| 7[315] | AATCCAAAAAAAAGGCTCCAAAA | |
| 10[583] | TGGCAAGTTTTTTGGGGTC | |
| 2[559] | GAATTATCCAATAACGATAGCTTAGATT | |
| 11[364] | GTCCACGCGCCACCTCACCGTTGAAACA | |
| 11[471] | TGTTCCAACGCTAACGAACAAGTCAGCAGGGAAGCGCATT | |
| 4[522] | ACCGCATTCCAACGGTATTCTAAGCGAGATATAGAAGGCT | |
| 7[380] | TCAAGCAGAACCACCACTCACTCAGGTAGCCCGGAATAGG | |
| 8[447] | ATTCTTTTCATAATCAAAATCAC | |
| 6[321] | AAATCCCCGAAACAATTCATGAGGAAGT | |
| 10[541] | CATTCTATCAGGGCGATGG | |
| 10[373] | TACCTGGTTTGCCCCAGCA | |
| 5[413] | AGAGTTTATACCAGTAGCACCTGAAACCATCGATA | |
| 9[105] | GTCCGTCCTGCAAGATCGTCGGATTCTCTTCGCATTGGACGA | |
| 11[63] | ATAGCTGTTTCCTGGAACGTCCATAACGCCGTAAA | |
| 11[177] | TGCGTACTAATAGTAGTTGAAATGCATATTTCAACGCAAG | |
| 8[702] | TTTTTAAAAACGCTCATGGAAATA | |
| 7[441] | TTGAAGCCCTTTTTAAGAAAAGT | |
| 11[525] | AGGGCGAAAAACCGATTTAACGTAGGGCAAATACC | |
| 2[475] | AAATAGGTAATTTACAAATAAGAAACGA | |
| 9[525] | TTTAGCAAACGCCACAATATAACTATATTCCCTTATAAATGG | |
| 7[399] | TATTGCCTTTAGCGTCAGACTGT | |
| 0[389] | GAATTGTAGCCAGAATGGATCAGAGCAAATCCT | |
| 2[307] | TTCCATTGACCCAAAGAGGCTTTGAGGA | |
| 7[651] | TAAGTAGAAGAACTCAAACTATCG | |
| 2[517] | ACGCGTCGGCTGTAAGACGACGACAATA | |
| 7[483] | GTTTACCGCGCCCAATAGCAAGC | |
| 2[55] | TTCGCCATAAACTCTGGAGGTGTCCAGC | |
| 10[48] | GCAGCACTTTGCTCTGAGCCGGGTCACTGTTGCCCTGCGGCTTTTT | |
| 6[657] | TGCCTGAACAGCAAATGAATGCGCGAACT | |
| 3[163] | TAAAGAGGCAAAATATTTTATAA | |
| 0[538] | TTAGGTTGGGTTATAGATAAGTC | |
| 4[419] | GCAGCACCGTAAGTGCCCGTATA | |
| 8[214] | CAAATGGTTCAGAAGAACGAGTAGAT | |
| 3[415] | GTTTATGTCACATGGGAATCCAC | |
| 0[641] | CCGAACCCCCTAAAACATCGACCAGTTTAGAGC | |
| 8[321] | CCGAACGGTGTACAGACCAGGCG | |
| 3[457] | CAATCCAAAATACTGAACAGTAG | |
| 6[195] | TGCAACTCAAAAGGCCGTACCAAAAACA | |
| 0[95] | CCGGAAGACGTACAGCGCCGCGATTACAATTCC | |
| 11[399] | GTTTGATGGTGGTTCAGAACCCCGCCTCACAGAAT | |
| 11[25] | TTTTTCCGGGTACCGAGCTCGAATTCGTAATCTGGTCA | |
| 0[53] | CGGTAGTACTCAATCCGCTGCTGGTCATGGTC | |
| 8[256] | AAAATTCCATTCAGGCTTTTGCAAAAGAAGTCA | |
| 3[266] | AACTTTAATCATGGGTAGCAACGGCTACGACAGCAACTAAAA | |
| 10[247] | AATAACGCGCGGGGAGAGGCGGTT | |
| 0[251] | TGGGAAGAAAAATCTACGTGCGTTTTAATT | |
| 0[263] | CAGTCTTGATTTTAAGAAC | |
| 8[286] | GACCTTCATTTTGCCAGAGGGGGTAATAGT | |
| 7[296] | AGACGTCGTCACCCTCAGACCTGCTC | |
| 4[461] | AAGAAACAATGACCGGAAACGTC | biotin labeled |
| 4[83] | GTACATCGACATCGTTAACGGCA | biotin labeled |
| 5[665] | ATACCACCATCAGTGAGGCCAAACCGTTGTAGCAA | biotin labeled |
| 4[251] | AACGCCAAAAGGCGGATGGCTTA | biotin labeled |
| 5[119] | CATAATATTCCGTAATGGGATCCGTGCATCTGCCA | external labeling |
| 3[98] | GGATAACCTCACAATTTTTGTTA | external labeling |
| 4[125] | GTTTGAGGGGACCTCATTTGCCG | external labeling |
| 4[144] | CGTAAAGGTCACGAAACCAGGCAATAGCACCGCTTCTGGT | external labeling |
| 0[137] | CATCAGCGTCTGGCCTTCCACAGGAACCTGGGG | external labeling |
| 10[121] | GGGCCGGAAGCATAAAGTG | external labeling |
| 11[135] | TAAAGGATTGTATAAGCGCACAAACGACATTAAATGTGAG | external labeling |
| 7[128] | TTCCGAATTGTAAACGTGTCGCCAGCATCGGTGCGGGCCT | external labeling |
| 3[140] | CAATAGGAACGCAAATTAAGCAA | external labeling |
| 7[105] | GAAAGATCGCACTCCAGCCAGCT | external labeling |
| 0[160] | GATAAAAATTTTTAGCCAGCTTT | external labeling |
| 8[153] | TCAGGCTGCGCAACTGTTGGGAA | external labeling |
| 0[118] | CGAGTAACAACCGTTTACCAGTC | external labeling |
| 2[139] | TTCGCGGATTGATTGCTCATTTTTTAAC | external labeling |
| 10[293] | ACCGGATGTTTTTCTTTTCACCA | 5’ ATTO 647N |
| 10[279] | ACCCAAATGGCAAAAGAATACTCGGAACAGAATCC | 5’ ATTO 647N |
| 10[286] | ATTCATTAGAGTAATCTTGACGCTGGCT | 5’ ATTO 647N |
| 10[265] | AACAAAGCTGCTGTAACAACAAGGACGT | 5’ ATTO 647N |
| 10[272] | TCAACGTTGCGTATTGGGCGCCAGGGTG | 5’ ATTO 647N |

## References

1. Derr, N. D. *et al.* Tug-of-war in motor protein ensembles revealed with a programmable DNA origami scaffold. *Science* **338,** 662–665 (2012).
2. Nickels, P. C. *et al.* Molecular force spectroscopy with a DNA origami-based nanoscopic force clamp. *Science* **354,** 305–307 (2016).
3. Vogelsang, J. *et al.* A reducing and oxidizing system minimizes photobleaching and blinking of fluorescent dyes. *Angew. Chem. Int. Ed.* **47,** 5465–5469 (2008).
4. Holzmeister, P., Wünsch, B., Gietl, A. & Tinnefeld, P. Single-molecule photophysics of dark quenchers as non-fluorescent FRET acceptors. *Photochem. Photobiol. Sci.* **13,** 853–858 (2014).
5. Weston, K. D. *et al.* Measuring the Number of Independent Emitters in Single-Molecule Fluorescence Images and Trajectories Using Coincident Photons. *Anal. Chem.* **74,** 5342–5349 (2002).
6. Stangl, T. *et al.* Mesoscopic quantum emitters from deterministic aggregates of conjugated polymers. *Proc. Natl. Acad. Sci.* **112**, E5560-E5566 (2015).
7. Eder, T. *et al.* Switching between H- and J-type electronic coupling in single conjugated polymer aggregates. *Nat. Commun.* **8**, 1641 (2017).
8. Morawska, P. O. *et al.* Side-Chain Influence on the Mass Density and Refractive Index of Polyfluorenes and Star-Shaped Oligofluorene Truxenes. *J. Phys. Chem. C* **119**, 22102-22107 (2015).
9. Manzano, D. A short introduction to the Lindblad master equation. *AIP Advances.* **10**, 025106 (2020).
10. Fischer, K. A. *et al.* Dynamical modeling of pulsed two-photon interference. *New J. Phys.* **18**, 113053 (2016).